\documentstyle[preprint,tighten,prc,aps]{revtex}

\begin{document}

\preprint{INT \#DOE/ER/40561-245-INT96-00-114}
\draft

\def\bra#1{{\langle#1\vert}}
\def\ket#1{{\vert#1\rangle}}
\def\coeff#1#2{{\scriptstyle{#1\over #2}}}
\def\undertext#1{{$\underline{\hbox{#1}}$}}
\def\hcal#1{{\hbox{\cal #1}}}
\def\sst#1{{\scriptscriptstyle #1}}
\def\eexp#1{{\hbox{e}^{#1}}}
\def\rbra#1{{\langle #1 \vert\!\vert}}
\def\rket#1{{\vert\!\vert #1\rangle}}
\def\lsim{{ <\atop\sim}}
\def\gsim{{ >\atop\sim}}
\def\nubar{{\bar\nu}}
\def\psibar{{\bar\psi}}
\def\Gmu{{G_\mu}}
\def\alr{{A_\sst{LR}}}
\def\wpv{{W^\sst{PV}}}
\def\evec{{\vec e}}
\def\notq{{\not\! q}}
\def\notk{{\not\! k}}
\def\notp{{\not\! p}}
\def\notpp{{\not\! p'}}
\def\notder{{\not\! \partial}}
\def\notcder{{\not\!\! D}}
\def\notA{{\not\!\! A}}
\def\notv{{\not\!\! v}}
\def\Jem{{J_\mu^{em}}}
\def\Jana{{J_{\mu 5}^{anapole}}}
\def\nue{{\nu_e}}
\def\mn{{m_\sst{N}}}
\def\mns{{m^2_\sst{N}}}
\def\me{{m_e}}
\def\mes{{m^2_e}}
\def\mq{{m_q}}
\def\mqs{{m_q^2}}
\def\mz{{M_\sst{Z}}}
\def\mzs{{M^2_\sst{Z}}}
\def\ubar{{\bar u}}
\def\dbar{{\bar d}}
\def\sbar{{\bar s}}
\def\qbar{{\bar q}}
\def\sstw{{\sin^2\theta_\sst{W}}}
\def\gv{{g_\sst{V}}}
\def\ga{{g_\sst{A}}}
\def\pv{{\vec p}}
\def\pvs{{{\vec p}^{\>2}}}
\def\ppv{{{\vec p}^{\>\prime}}}
\def\ppvs{{{\vec p}^{\>\prime\>2}}}
\def\qv{{\vec q}}
\def\qvs{{{\vec q}^{\>2}}}
\def\xv{{\vec x}}
\def\xpv{{{\vec x}^{\>\prime}}}
\def\yv{{\vec y}}
\def\tauv{{\vec\tau}}
\def\sigv{{\vec\sigma}}
\def\sst#1{{\scriptscriptstyle #1}}
\def\gpnn{{g_{\sst{NN}\pi}}}
\def\grnn{{g_{\sst{NN}\rho}}}
\def\gnnm{{g_\sst{NNM}}}
\def\hnnm{{h_\sst{NNM}}}

\def\xivz{{\xi_\sst{V}^{(0)}}}
\def\xivt{{\xi_\sst{V}^{(3)}}}
\def\xive{{\xi_\sst{V}^{(8)}}}
\def\xiaz{{\xi_\sst{A}^{(0)}}}
\def\xiat{{\xi_\sst{A}^{(3)}}}
\def\xiae{{\xi_\sst{A}^{(8)}}}
\def\xivtez{{\xi_\sst{V}^{T=0}}}
\def\xivteo{{\xi_\sst{V}^{T=1}}}
\def\xiatez{{\xi_\sst{A}^{T=0}}}
\def\xiateo{{\xi_\sst{A}^{T=1}}}
\def\xiva{{\xi_\sst{V,A}}}

\def\rvz{{R_\sst{V}^{(0)}}}
\def\rvt{{R_\sst{V}^{(3)}}}
\def\rve{{R_\sst{V}^{(8)}}}
\def\raz{{R_\sst{A}^{(0)}}}
\def\rat{{R_\sst{A}^{(3)}}}
\def\rae{{R_\sst{A}^{(8)}}}
\def\rvtez{{R_\sst{V}^{T=0}}}
\def\rvteo{{R_\sst{V}^{T=1}}}
\def\ratez{{R_\sst{A}^{T=0}}}
\def\rateo{{R_\sst{A}^{T=1}}}

\def\mro{{m_\rho}}
\def\mks{{m_\sst{K}^2}}
\def\Qhat{{\hat Q}}

\def\FOS{{F_1^{(s)}}}
\def\FTS{{F_2^{(s)}}}
\def\GAS{{G_\sst{A}^{(s)}}}
\def\GES{{G_\sst{E}^{(s)}}}
\def\GMS{{G_\sst{M}^{(s)}}}
\def\GATEZ{{G_\sst{A}^{\sst{T}=0}}}
\def\GATEO{{G_\sst{A}^{\sst{T}=1}}}
\def\mdax{{M_\sst{A}}}
\def\mustr{{\mu_s}}
\def\rsstr{{r^2_s}}
\def\rhostr{{\rho_s}}
\def\GEG{{G_\sst{E}^\gamma}}
\def\GEZ{{G_\sst{E}^\sst{Z}}}
\def\GMG{{G_\sst{M}^\gamma}}
\def\GMZ{{G_\sst{M}^\sst{Z}}}
\def\GEn{{G_\sst{E}^n}}
\def\GEp{{G_\sst{E}^p}}
\def\GMn{{G_\sst{M}^n}}
\def\GMp{{G_\sst{M}^p}}
\def\GAp{{G_\sst{A}^p}}
\def\GAn{{G_\sst{A}^n}}
\def\GA{{G_\sst{A}}}
\def\GETEZ{{G_\sst{E}^{\sst{T}=0}}}
\def\GETEO{{G_\sst{E}^{\sst{T}=1}}}
\def\GMTEZ{{G_\sst{M}^{\sst{T}=0}}}
\def\GMTEO{{G_\sst{M}^{\sst{T}=1}}}
\def\lamd{{\lambda_\sst{D}^\sst{V}}}
\def\lamn{{\lambda_n}}
\def\lams{{\lambda_\sst{E}^{(s)}}}
\def\bvz{{\beta_\sst{V}^0}}
\def\bvo{{\beta_\sst{V}^1}}
\def\Gdip{{G_\sst{D}^\sst{V}}}
\def\GdipA{{G_\sst{D}^\sst{A}}}

\def\RAp{{R_\sst{A}^p}}
\def\RAn{{R_\sst{A}^n}}
\def\RVp{{R_\sst{V}^p}}
\def\RVn{{R_\sst{V}^n}}
\def\rva{{R_\sst{V,A}}}

\def\PR#1{{{\em   Phys. Rev.} {\bf #1} }}
\def\PRC#1{{{\em   Phys. Rev.} {\bf C#1} }}
\def\PRD#1{{{\em   Phys. Rev.} {\bf D#1} }}
\def\PRL#1{{{\em   Phys. Rev. Lett.} {\bf #1} }}
\def\NPA#1{{{\em   Nucl. Phys.} {\bf A#1} }}
\def\NPB#1{{{\em   Nucl. Phys.} {\bf B#1} }}
\def\AoP#1{{{\em   Ann. of Phys.} {\bf #1} }}
\def\PRp#1{{{\em   Phys. Reports} {\bf #1} }}
\def\PLB#1{{{\em   Phys. Lett.} {\bf B#1} }}
\def\ZPA#1{{{\em   Z. f\"ur Phys.} {\bf A#1} }}
\def\ZPC#1{{{\em   Z. f\"ur Phys.} {\bf C#1} }}
\def\etal{{{\em   et al.}}}

\def\delalr{{{delta\alr\over\alr}}}

\title{Chiral Symmetry and the Nucleon's\\
Vector Strangeness Form Factors}

\author{M.J. Musolf$^{a,b}$\thanks{On leave from the Department of
Physics, Old Dominion University, Norfolk, VA 23529 USA} and 
Hiroshi Ito$^{c}$
}
\address{
$^a$ Continuous Electron Beam Accelerator Facility 
Newport News, VA 23606 USA \\
$^b$ Institute for Nuclear Theory, 
University of Washington, Seattle, WA 98195 USA\\
$^c$ Center for Nuclear Studies,
	Department of Physics, George Washington University \\
Washington, D.C. 20052 USA
}


\maketitle

\begin{abstract}

The nucleon's strange-quark vector current form factors are studied
from the perspective of chiral symmetry. It is argued that chiral
perturbation theory cannot yield a prediction for the strangeness radius
and magnetic moment. Arrival at definite predictions requires
the introduction of additional, model-dependent assumptions which go
beyond the framework of chiral perturbation theory.
A variety of such model predictions is surveyed,
and the credibility of each is evaluated.
The most plausible prediction appears in a model where the unknown
chiral counterterms are identified with $t$-channel vector meson exchange
amplitudes. The corresponding prediction for the mean square Dirac
strangeness radius is $\langle r_s^2\rangle = 0.24$ fm$^2$, which would
be observable in up-coming semileptonic determinations of
the nucleon's strangeness form factors.
\end{abstract}


\narrowtext

\section{Introduction}
\label{sec:intro}
\medskip
There has been considerable interest recently in the strange quark 
\lq\lq content" of the nucleon 
\cite{Mus94a,Mck89,Pit94,Bec91,Bei91,Fin91,Har93,Mck95,Lou89,Mus92,Gar92,Hor93,Bar96,Liu95,Lei96}. 
The reasons for this interest
are both theoretical and phenomenological. In the latter case, early analyses
of the pion-nucleon sigma term \cite{Che76} and later results for the nucleon's
inclusive, spin-dependent deep-inelastic structure functions 
\cite{Ash89,Ant93,Ade93,Ada94,Abe95}
suggested that a non-trivial fraction of the nucleon's mass and spin are
carried by the $s\bar s$ component of the sea. Subsequent analyses
of the sigma-term have reduced the value of $\bra{p}\sbar s\ket{p}/
\bra{p}\ubar u+\dbar d\ket{p}$, and therefore the strange-quark contribution
to $\mn$, by a factor of two \cite{Gas91}, 
while studies of SU(3)-breaking in the 
axial vector octet imply a theoretical uncertainty in the value of $\Delta s$,
the strange quark contribution to the nucleon's spin extracted from 
deep inelastic scattering (DIS)
measurements, sufficiently large to make the bounds on $\Delta s$ consistent
with zero \cite{Jen91,Dai95,Ehr94,Lic95}. 
Nevertheless, the early analyses of the sigma term and 
polarized DIS
results have motivated proposals to measure another strange-quark
observable, $\bra{p}\sbar\gamma_\mu s\ket{p}$. Indeed, 
several low- and medium-energy parity-violating
electron scattering experiments are either underway or planned at
MIT-Bates \cite{Mck89,Pit94}, CEBAF \cite{Bec91,Bei91,Fin91} 
and Mainz \cite{Har93} with the goal of measuring
the two form factors which parameterize the nucleon's strange quark vector
current, $\GES$ and $\GMS$.  

Theoretically, strange-quarks are interesting because they don't 
appear explicitly in most quark model descriptions of the nucleon.
Although the quark model provides a useful intuitive picture of the nucleon's
substructure and has seen considerable success in accounting for a wide range of
properties of the low-lying hadrons \cite{Cap95}, 
one knows that there is more to
the nucleon than the three constituent quarks. In particular, processes
such as DIS and Drell-Yan have provided considerable insight regarding
the important role played by the $q\bar q$ and gluon sea when the nucleon
interacts at high energies \cite{Mck95}. 
Almost no information exists, however,  
regarding the low-energy manifestations of the sea.
Because strange quarks constitute purely sea degrees of
freedom, low- and intermediate-energy determinations
of strange quark matrix elements offer a new window on the 
\lq\lq low-energy" structure of the
nucleon which goes beyond the description provided by the quark model. In
particular, the weak neutral current
scattering experiments mentioned above should set bounds
on the spatial polarization of the $s\sbar$ sea \cite{Bec91,Fin91}, 
its contributions to the
nucleon magnetic moment \cite{Mck89,Pit94,Bec91} and spin \cite{Lou89}, 
and its role in the nuclear response at
moderate momentum transfer \cite{Bei91}. 

	One has seen considerable progress over the past few years in
clarifying the interpretation of neutral current
observables in terms of strangeness
matrix elements \cite{Mus94a,Mus92,Gar92,Hor93,Bar96}. 
The situation regarding theoretical predictions for
these matrix elements is less advanced. Ideally, one would hope to draw
inferences from the deep inelastic data on $s$- and $\sbar$-distributions
\cite{ssb} for 
elastic vector and axial vector strangeness matrix elements. However,
the high-energy data provide light-cone momentum distribution functions,
and one does not know at present how to translate this information into the
spin and spatial nucleon wavefunctions as needed to compute charge radii,
magnetic moments, {\em  etc.} \cite{Neg}. 
Similarly, one might hope for first-principles
microscopic predictions using lattice QCD. To date, lattice results for
the strangeness axial charge \cite{Liu95} and strangeness magnetic moment 
\cite{Lei96}
have been obtained in the quenched approximation, and one anticipates
a refinement of these results as lattice methods continue to advance. 
In the absence of definitive lattice calculations -- and with an eye
toward understanding the mechanisms which govern the scale of nucleon
strangeness -- a variety of model calculations have been performed.  
The latter have yielded a wide array of predictions for
strangeness matrix elements which vary in both magnitude and sign 
\cite{Jaf89,Mus94b,Koe92,For94,Coh93,Par91,Hon93,Pha94,Ito95,Gei96}.
While one might argue {\em ad nauseum} about the relative merits
of different models,  there is no compelling reason to take any particular model
calculation as definitive. 

	In an effort to add some clarity to this situation,
we discuss in this paper the implications for nucleon strangeness vector
current matrix elements of one of the underlying,
approximate symmetries of QCD: chiral
symmetry. The use of chiral symmetry, in the guise of chiral perturbation
theory (CHPT), has proven highly effective in predicting and interpreting
a wide variety of low-energy observables \cite{Don92,Mei92}. 
The essential strategy of CHPT
is to exploit the approximate $\hbox{SU(3)}_L\times\hbox{SU(3)}_R$ chiral
symmetry of QCD for the three lightest flavors to relate one set of 
observables to another (accounting for loop effects), or to draw on 
one set of measured quantities to predict another. 
This approach has recently been employed to analyze of baryon octet and decuplet
magnetic moments \cite{Jen93,But94,Ban96}
and the nucleon's isovector charge 
radius \cite{Coh95}. As we illustrate below,
this strategy breaks down in the flavor-singlet channel, rendering CHPT
un-predictive for the nucleon's strangeness matrix elements. The reason
is that the coefficients of the relevant flavor-singlet operators in the
chiral Lagrangian, which contain
information on short-distance hadronic effects, cannot be determined from
existing measurements by using chiral symmetry. 
Although the leading, non-analytic
long-distance (loop) contributions are calcuable (${\cal O}(\sqrt{m_s})$ for the
strangeness magnetic moment and ${\cal O}(\ln m_s)$ for the strangeness
radius), one has no reason to assume that they are numerically more 
important than the unknown analytic terms arising at the same or lower order 
from the chiral Lagrangian. The only rigorous way to determine these 
unknown analytic contributions is to measure the very quantity one would
like to predict: the nucleon's flavor-singlet current matrix element.

Consequently,
if one wishes to make any predictions at all, one must invoke additional --
and therefore model-dependent -- assumptions. We illustrate this next line
of defense in three forms: (a) a \lq\lq resonance saturation" model
in which the unknown constants arising
in chiral perturbation theory are determined by the $t$-channel exchange of
vector mesons; (b) a class of models in which the nucleon's \lq\lq kaon
cloud" is assumed to dominate the strangeness form factors; and (c)
constituent chiral quark models in which nucleon's strangeness matrix elements
arise from the strangeness content of the constituent $U$- and $D$-quarks.
For each of these approaches, we present new calculations and compare
them with calculations discussed elsewhere in the literature.
The corresponding results are 
unabashedly model-dependent and, therefore, not strong. 
We give them mainly to illustrate the outer limits to which one might go in 
employing chiral symmetry to compute $\GES$ and $\GMS$. 
Although there exist additional chiral
model approaches not considered in detail here, we believe that the three 
which we discuss are sufficiently representative so as to illustrate the 
breadth of predictions permitted by chiral symmetry. Among these predictions,
there does appear to be one 
having a greater degree of credibility than others: the value of the
Dirac mean square strangeness radius arising from the vector meson exchange,
or resonance saturation,
model. Nevertheless, even in this case, the logic is not airtight. A more
detailed analysis of the strangeness and isoscalar electromagnetic form
factors within the framework of dispersion relations
could reveal important contributions not included in the resonance
saturation approach.

	We organize our discussion of these points as follows. In 
Section II
we review the effective low-energy chiral Lagrangians which describe the
interaction of pseudoscalar mesons with baryons or quarks. In Section III
we employ this formalism to compute the nucleon's strange-quark vector
current form factors, introducing model assumptions as necessary and
evaluating their credibility.
Section IV gives the 
results of these calculations and a discussion of
their meaningfulness. Section V summarizes our conclusions.

\def\pitt{{\tilde\Pi}}
\def\lamchi{{\Lambda_\chi}}
\def\lamchis{{\Lambda_\chi^2}}
\def\bbar{{\bar B}}

\section{Chiral Lagrangians}
\label{sec:secII}

In the low-energy world of three-flavor QCD, the QCD Lagrangian manifests
an approximate SU(3$)_L\times$SU(3$)_R$ chiral symmetry. This symmetry is
explicitly broken by the small current quark masses. In addition, spontaneous
symmetry-breaking SU(3$)_L\times$SU(3$)_R\to$SU(3$)_V$ implies the existence of
eight massless (assuming $\mq=0$) Goldstone modes and an axial vector 
condensate. One identifies the latter with the pion decay constant $f_\pi
\approx 93$ MeV and the former with the lowest-lying octet of pseudoscalar
mesons. The Goldstone bosons are conveniently described by a field $\Sigma$,
given by
\begin{equation}
\Sigma = \hbox{exp}\left[2i\pitt/f\right]\ \ \ ,
\label{eq:sig}
\end{equation}
where $f\equiv f_\pi$ and 
\begin{equation}
\pitt={1\over 2}\sum_{a=1}^{8} \lambda_a\phi_a\ \ \ ,
\label{eq:moct}
\end{equation}
with the $\lambda_a$ being the eight Gell-Mann matrices and the $\phi_a$
being the pseudoscalar meson fields \cite{Geo84,Man84}. 
The Lagrangian which describes the
pseudoscalar kinetic energies and self interactions is given by
\begin{equation}
{\cal L}={f^2\over 4}\hbox{ Tr}\left(\partial^\mu\Sigma^{\dag}
   \partial_\mu\Sigma\right)+{f^2\over 2}\left[\hbox{ Tr}\left(\Sigma\mu M
   \right) + \hbox{ h.c.}\right]\ \ \ ,
\label{eq:lpi}
\end{equation}
where $M=\hbox{ diag}[m_u, m_d, m_s]$ is just the QCD current quark mass matrix 
which explicitly breaks the residual SU(3$)_V$ symmetry
and $\mu$ is a parameter which relates the quark masses to quadratic forms
in the pseudscalar masses (hence, $m_{\pi, K}$ is of order $\sqrt{m_q}$).
The Lagrangian in Eq.~(\ref{eq:lpi}) actually constitutes
the leading term in an expansion in powers of $p/\lamchi$ and $\mu M/\lamchi$,
where $p$ denotes the momentum of a low-energy pseudoscalar meson and 
$\lamchi\approx 4\pi f\approx 1$ GeV is the scale of chiral symmetry breaking.
For purposes of the present study, retention of higher-order terms in 
the chiral expansion of the purely mesonic sector is not necessary.

	Interactions between the Goldstone bosons and matter fields are
conveniently described by first introducing vector and axial vector currents
\begin{eqnarray}
V_\mu & \equiv & {1\over 2}(\xi^{\dag}\partial_\mu\xi+\xi\partial_\mu
	\xi^{\dag})
\label{eq:vec}\\
	A_\mu & \equiv & {i\over 2}(\xi^{\dag}\partial_\mu\xi-\xi\partial_\mu
	\xi^{\dag})\ \ \ , 
\label{eq:axi}
\end{eqnarray}
where $\Sigma=\xi^2$. One may now proceed to construct a chiral
Lagrangian for fermions. The simplest case involves the effective, consituent
quarks of the quark model. Letting
\begin{equation}
\psi=\left(\matrix{U\cr D\cr S\cr}\right)
\end{equation}
denote the triplet of light-quark fields, one has for the leading term in the
chiral expansion
\begin{equation}
{\cal L}_Q = \psibar\left( i\notcder - m\right)\psi + \ga\psibar
	\notA\gamma_5\psi\ \ \ ,
\label{eq:lquark}
\end{equation}
where
\begin{equation}
D_\mu\equiv\partial_\mu+V_\mu
\label{eq:cov}
\end{equation}
is a chiral covariant derivative and $\ga$ is a constant which governs the
strength of the interaction between quarks and odd numbers of pseudoscalar
mesons (the last term in Eq.~(\ref{eq:lquark})). 
The term involving $m$ gives the 
constituent quark masses\footnote{not to be confused with the
current quark mass matrix $M$} 
in the limit of good SU(3$)_V$ symmetry. Higher-order
terms in the chiral expansion include those which break the degeneracy between
the constituent quarks. In the chiral quark model calculation we discuss
below, we allow for mass splittings among the constituent quarks, although
we will not show the SU(3$)_V$ symmetry-breaking terms explicitly. The
higher order terms in the chiral expansion relevant to strangeness vector
current matrix elements will be introduced below.

	In the case of meson-baryon interactions, more thought is required
when writing down an effective chiral Lagrangian. 
In the most na\"\i ve approach,
one assigns baryons to the appropriate SU(3) multiplet and constructs objects
using this multiplet, derivatives, $V_\mu$ and $A_\mu$ which transform as 
SU(3$)_L\times$SU(3$)_R$ singlets. For the lowest-lying octet of baryons,
one has the matrix representation
\begin{equation}
B\equiv {1\over\sqrt{2}}\sum_{a=1}^8 \lambda_a\psi_a\ \ \ ,
\label{eq:bact}
\end{equation}
where the $\psi_a$ are the octet baryon fields. The na\"\i ve, leading-order
baryon Lagrangian is, then,
\begin{equation}
{\cal L}_B=\hbox{ Tr}\left[{\bar B}(i\notcder-m_B)B\right]+ D
\hbox{ Tr}\left(\bbar\gamma_\mu\gamma_5\{A^\mu, B\}\right)+F\hbox{ Tr}
\left(\bbar\gamma_\mu\gamma_5[A^\mu, B]\right)\ \ ,
\label{eq:lbar}
\end{equation}
where, in this case, the action of the chiral covariant derivative on the
baryon fields is given by
\begin{equation}
D^\mu B = \partial^\mu B + [V^\mu, B]\ \ \ ,
\label{eq:dbar}
\end{equation}
and where $m_B$ gives the octet baryon masses in the limit of good
SU(3$)_V$ symmetry. The constants $D$ and $F$ are the usual SU(3) 
reduced matrix elements. 

	As in the case of the quark chiral Lagrangian, one could formally
write down corrections to ${\cal L}_B$ as a series in powers of $p/\lamchi$
and $m_B/\lamchi$. Such an expansion would not be convergent, however,
since numerically one has $m_B\sim\lamchi$. This situation contrasts with
that of chiral quarks whose constituent masses are well below the scale of
chiral symmetry breaking. Consequently, one has no reason to believe that
higher-order terms in the chiral expansion are less important than the terms
given in Eq.~(\ref{eq:lbar}). Jenkins and Manohar \cite{Jen91} 
developed an approach
to circumvent this difficutly with the baryon chiral Lagrangian. The idea
is to approximate the baryons as very heavy static fields whose spectrum is
characterized by states of good velocity, $v_\mu=p_\mu/m_B$. One
correspondingly redefines the baryon fields by rotating away the frequency
dependence associated with the heavy mass:
\begin{equation}
B_v(x)\equiv \hbox{ exp}(i m_B\ \notv v\cdot x) B(x)\ \ \ .
\label{eq:bheav}
\end{equation}
The new baryon Lagrangian, specified now for baryons of given velocity,
has a similar form to that appearing in Eq.~(\ref{eq:lbar}) 
but with no baryon mass term.
As a result, one can now perform a chiral expansion in powers of external
momenta and baryon mass {\em splittings} divided by $\lamchi$ without
encountering the problematic $m_B/\lamchi$ terms. Before writing this 
Lagrangian, we follow Ref. \cite{Jen91} and make use of the spin operators 
$S_v^\lambda$ which satisfy the relations
\begin{eqnarray}
v\cdot S =0;&\ \ S_v^2 B_v& = -(3/4)B_v;\ \ \{S_v^\alpha, S_v^\beta\}= 
(1/2)(v^\alpha v^\beta - g^{\alpha\beta}); 
\nonumber \\
\nonumber \\
\left[ S_v^\alpha, S_v^\beta \right]
=&i\epsilon^{\alpha\beta\mu\nu} & v_\mu S_{v\nu}\ \ \ .
\label{eq:identa}
\end{eqnarray}
In addition, we use the relations
\begin{equation}
\bbar_v\gamma^\mu B_v = v^\mu \bbar_v B_v;\quad \bbar_v\gamma^\mu\gamma_5
B_v = 2 \bbar_v S_v^\mu B_v;\quad \bbar_v\sigma^{\mu\nu} B_v = 
2\epsilon^{\mu\nu\alpha\beta} v_\alpha \bbar_v S_{v\beta} B_v\ .
\label{eq:identb}
\end{equation}
Using these identities, the leading-order heavy baryon chiral Lagrangian is
\begin{equation}
{\cal L}_v=i\hbox{ Tr}\left({\bar B}_vv\cdot D\ B_v\right)+ 2D
\hbox{ Tr}\left(\bbar_v S_v^\mu\{A_\mu, B_v\}\right)+2F\hbox{ Tr}
\left(\bbar_v S_v^\mu[A_\mu, B_v]\right)\ .
\label{eq:lheav}
\end{equation}
The first corrections to ${\cal L}_v$ contain one more power of the
chiral covariant derivative, quark mass matrix $M$, or axial vector 
$A_\mu$.

In what follows, we compute strange quark vector currents of non-strange
chiral quarks and non-strange baryons arising from kaon loops. To that end,
it is useful to work instead with the baryon number current $J_\mu^\sst{B}$
and to introduce a vector current source $Z^\mu$ which couples to
$J_\mu^\sst{B}$ via the minimal substitution $\partial^\mu\to\partial^\mu
+i{\hat Q}_B Z^\mu$, where ${\hat Q}_B$ is the baryon number 
operator.\footnote{The full chiral structure of the charge operator
is given in Ref.~\cite{Jen93}. For the present purpose, the inclusion
of the full structure is not necessary.}
Taking the first functional derivative with respect to $Z^\mu$
of the generating functional yields
$n$-point functions with a single $J_\mu^\sst{B}$ insertion.
The strange quark current is related in a straightforward manner to
$J_\mu^\sst{B}$ and the isoscalar EM current (see Eqs.~(\ref{eq:jemo}-
\ref{eq:scur}) below).
In practice, it is simpler to 
compute the strangeness charge of each particle appearing in a Feynman diagram,
insert the appropriate Lorentz structure for a vector current, 
and evaluate the resulting contribution to the strangeness matrix element.
From a formal standpoint, however, the use of the baryon number current and
of the source $Z^\mu$ provides an efficient means for keeping track of the
flavor content and chiral order associated with higher moments (mean square
radius, magnetic moment, {\em etc.}) of various currents.

\section{Strange Quark Matrix Elements}
\label{sec:secIII}

With the formalism of Section II in hand, it is straightforward
to compute nucleon matrix elements of the strange-quark vector current,
$\bra{p'} \sbar\gamma_\mu s\ket{p}$. This matrix element can be parameterized
in terms of two form factors, $\FOS$ and $\FTS$:
\begin{equation}
\bra{p'}\sbar\gamma_\mu s\ket{p} =\ubar(p')\left[\FOS\gamma_\mu
+i{\FTS\over 2\mn}\sigma_{\mu\nu}Q^\nu\right] u(p)\ \ \ ,
\label{eq:smat}
\end{equation}
where $u(p)$ denotes a nucleon spinor and $Q=p'-p$ is the momentum transfer
to the nucleon. When working in the heavy baryon formalism, the corresponding
Lorentz structures are obtained from Eq.~(\ref{eq:smat}) 
by the use of the relations in Eq.~(\ref{eq:identa}). 
For on-shell nucleons, the form factors are functions of
$Q^2=q_0^2-|\qv|^2$, where $Q^\mu=(q_0, \qv)$.
In what follows, we work with the so-called Sachs electric and magnetic form
factors \cite{Sac62}, defined as
\begin{eqnarray}
\GES&=&\FOS-\tau\FTS 
\label{eq:geff} \\
	 \GMS&=&\FOS+\FTS\ \ \ , 
\label{eq:gmff} 
\end{eqnarray}
where $\tau\equiv -Q^2/4\mns$. At $Q^2=0$, the Sachs electric form factor
gives the net strangeness of the nucleon, which is zero. At small momentum
transfer, the scale of this form factor is governed by the first derivative
with respect to $Q^2$, which defines the mean square \lq\lq strangeness 
radius". We work with a dimensionless version of this quantity, 
$\rho^s_\sst{S}$, defined as
\begin{equation}
\rho^s_\sst{S}={d\GES\over d\tau}\Big\vert_{\tau=0} = -{2\over 3}\mns\langle
r_s^2\rangle^{\sst{S}}\ \ \ ,
\label{eq:rhos}
\end{equation}
where $\langle r_s^2\rangle^{\sst{S}}$ is 
the dimensionful Sachs strangeness radius and where the superscript 
\lq\lq S" denotes the Sachs, as distinct from the Dirac, radius.
There exists no symmetry principle which constrains the strangeness magnetic
moment, $\GMS(0)=\mu^s$. Note that since $\GES(0)=0$ one has $\mu^s=
\kappa^s$.
In discussing the implications of chiral symmetry
for $\bra{p'}\sbar\gamma_\mu s\ket{p}$, we will be concerned primarily
with these two parameters, $\rho^s$ and $\mu^s$.

\medskip
\centerline{\bf A. Heavy Baryon Chiral Perturbation Theory}
\medskip
In terms of chiral counting, the strangeness magnetic moment and radius,
like the corresponding electromagnetic quantities, appear, respectively,
as order $1/\lamchi$ and $1/\lamchi^2$ corrections to the leading order
heavy baryon Lagrangian given in Eq.~(\ref{eq:lheav}). 
In discussing these corrections,
it is convenient to rewrite the strangeness vector current in terms of the
electromagnetic and baryon number currents:
\begin{eqnarray}
J_\mu^\sst{EM}(T=1)&=&V_\mu^{(3)}
\label{eq:jemo} \\
	 J_\mu^\sst{EM}(T=0)&=&(1/\sqrt{3})V_\mu^{(8)}
\label{eq:jemz} \\
	 J_\mu^\sst{B}&=&V_\mu^{(0)}\ \ \ ,
\label{eq:jbar}
\end{eqnarray}
where the $T=1$ and $T=0$ designations indicate the isovector and
isoscalar elecromagnetic currents, where \lq\lq B" denotes the
baryon number current, and where
\begin{equation}
V_\mu^{(a)}={\bar q}{\lambda^a\over 2}\gamma_\mu q\qquad\qquad q=
\pmatrix{u\cr d\cr s}\ \ \ .
\end{equation}
Here, the $\lambda^a, a=1,\ldots 8$ are the usual Gell-Mann matrices,
$\lambda^0=\coeff{2}{3}\hbox{\bf I}$, and $q$ gives the triplet of QCD
quark fields. In terms of the currents in Eq.~(\ref{eq:jemo}-\ref{eq:jbar}) one has
\begin{equation}
\sbar\gamma_\mu s= J_\mu^\sst{B}-2 J_\mu^\sst{EM}(T=0)\ \ \ .
\label{eq:scur}
\end{equation}
With these definitions one may write down the higher-order heavy baryon
Lagrangians corresponding to the EM and baryon number magnetic moments and
charge radii:
\begin{eqnarray}
\Delta{\cal L}_\sst{EM}^{\sst{T=1}}&=&{e\over\lamchi}
\epsilon_{\mu\nu\alpha\beta}v^\alpha\left\{b_+\hbox{ Tr}\left(
\bbar_v S_v^\beta \{\lambda^3,B_v\}\right)+b_-\hbox{ Tr}\left(
\bbar_v S_v^\beta [\lambda^3,B_v]\right)\right\}F^{\mu\nu}
\label{eq:ctem1} \\
&- &{e\over\lamchis}\left\{c_+\hbox{ Tr}\left(
\bbar_v \{\lambda^3, B_v\}\right)+c_-\hbox{ Tr}\left(
\bbar_v [\lambda^3, B_v]\right)\right\}v_\mu\partial_\lambda F^{\mu\lambda}
\nonumber\\
&&\nonumber\\
\Delta{\cal L}_\sst{EM}^{\sst{T=0}}&=&{e\over\lamchi}{1\over\sqrt{3}}
\epsilon_{\mu\nu\alpha\beta}v^\alpha\left\{b_+\hbox{ Tr}\left(
\bbar_v S_v^\beta \{\lambda^8,B_v\}\right)+b_-\hbox{ Tr}\left(
\bbar_v S_v^\beta [\lambda^8,B_v]\right)\right\}F^{\mu\nu}
\label{eq:ctem2} \\
&- &{e\over\lamchis}{1\over\sqrt{3}}\left\{c_+\hbox{ Tr}\left(
\bbar_v \{\lambda^8, B_v\}\right)+c_-\hbox{ Tr}\left(
\bbar_v [\lambda^8, B_v]\right)\right\}v_\mu\partial_\lambda F^{\mu\lambda}
\nonumber \\
&&\nonumber \\
\Delta{\cal L}_\sst{B}&=&{b_0\over\lamchi}
\epsilon_{\mu\nu\alpha\beta}v^\alpha\hbox{ Tr}\left(
\bbar_v S_v^\beta B_v\right)Z^{\mu\nu}
\label{eq:ctbar}\\
&-&{c_0\over\lamchis}\hbox{ Tr}\left(\bbar_v B_v\right)
v_\mu\partial_\lambda Z^{\mu\lambda}\ \ \ ,
\nonumber
\end{eqnarray}
where $F^{\mu\nu}$ is just the ordinary EM field strength tensor,
$Z^{\mu\nu}$ is the analogous quantity involving the source $Z^\mu$ coupling
to baryon number, and $e$ is the proton's EM charge.
In each $\Delta{\cal L}$, the terms of order $1/\lamchi$
contribute to the anomalous magnetic moment and those of order $1/\lamchis$
enter the charge radius \cite{Expa}. For a given
baryon, the magnetic moment and mean square radius will contain a contribution
from $\Delta{\cal L}$ and a contribution from loops (non-analytic in
$m_q$), as in Fig. 1:
\begin{eqnarray}
\kappa^a&=&\kappa^a_\sst{LOOP}+ \left({2m_B\over\lamchi}\right)b^a
\label{eq:kapa} \\
	 \rho^a_D&=&\rho^a_\sst{LOOP}-\left({2 m_B\over\lamchi}\right)^2 c^a
	\ \ \ , 
\label{eq:rhoa}
\end{eqnarray}
where $\kappa^a= F_2^{(a)}(0)$ is the anomalous magnetic moment, 
\lq\lq a" denotes
the corresponding flavor channel ($T=0,1$, $s$, SU(3) singlet), 
and the subscript 
\lq\lq D" indicates the slope of the Dirac form factor 
$(F_1)$ at $\tau=0$. In the
case of the EM moments, the quantities $b^a$ and $c^a$ contain appropriate
linear combinations of $b_{\pm}$ and $c_{\pm}$ as determined from the
traces appearing in Eqs. (\ref{eq:ctem1}-\ref{eq:ctem2}).
Using the heavy baryon formalism outlined above, we compute
$\kappa^a_\sst{LOOP}$ and $\rho^a_\sst{LOOP}$ employing dimensional
regularization. For the strangeness moments, we find
\begin{eqnarray}
	\kappa^s_\sst{LOOP}&=&(2\pi)\left[
	\left({3F+D\over\sqrt{6}}\right)^2+{3\over 2}(D-F)^2\right] 
	{\mn\over\lamchi}{m_K\over\lamchi} 
\label{eq:ksloop}\\
	\rho^s_\sst{LOOP}&=&\left({\mn\over\lamchi}\right)^2\left\{1+
	{5\over 3}\left[\left({3F+D\over\sqrt{6}}\right)^2
	+{3\over 2}(D-F)^2\right]\right\}
	\left[{\cal C}_\infty -\ln{m_K^2\over\mu^2}\right]
	\ \ ,
\label{eq:rsloop} 
\end{eqnarray}
where ${\cal C}_\infty=1/\varepsilon-\gamma+\ln 4\pi$ with $\varepsilon
=(4-d)/2$ and $d$ being the number of dimensions. One finds analogous 
expressions for the isovector ($\lambda^3$) and isoscalar 
($\lambda^8/\sqrt{3}$) components of the EM moments \cite{Jen93,Mustbp}. 
The constant $c^s$ appearing in Eq.~(\ref{eq:rhoa})
contains the appropriate dependence on ${\cal C}_\infty$ to cancel the
pole term in $\rho^s_\sst{LOOP}$. The scale $\mu$ denotes the scale at which
the subtraction of the pole term is carried out. The remaining finite parts
of $(\kappa^s_\sst{LOOP}, \rho^s_\sst{LOOP})$ and of $(b^s, c^s)$ determine
the value of the anomalous magnetic moment and mean square Dirac radius. 
Using Eqs.~(\ref{eq:scur}-\ref{eq:ctbar}), 
one can express the \lq\lq low-energy" constants $(b^s, c^s)$ 
in terms of the corresponding quantities for the baryon and 
EM currents\footnote{Henceforth, the cancellation of the ${\cal C}_\infty$
will be understood and $(b^a, c^a)$ will denote the finite remainders of
the counterterms.}:
\begin{eqnarray}
	  b^s&=&b_0-2[b_- - (b_+/3)]
\label{eq:bcts}\\
	  c^s&=&c_0-2[c_- - (c_+/3)]\ \ \ .
\label{eq:ccts}
\end{eqnarray}

In the case of the EM moments, the $(b_\pm, c_\pm)$ are fit to known EM moments
in the baryon octet. One may then employ Eqs.~(\ref{eq:ctem1}-\ref{eq:rhoa})
and the loop contributions
to predict the moments of other baryons within the octet. This approach
reflects the basic strategy of chiral perturbation theory: 
rely on chiral symmetry to relate one set of quantities (known EM moments)
to another (those one wishes to predict), modulo loop corrections (a
consequence of spontaneous chiral symmetry breaking). 
A simple fit to the nucleon EM moments alone gives
$b_+\approx 1.4$, $b_-\approx 0.9$, $c_+\approx -1.9$, 
$c_-\approx 0.9$ \cite{Expb}. 

As one would
expect on general grounds, these constants are of order unity. In the case
of the nucleon EM anomalous magnetic moments, the contributions from the
$b_\pm$ and the loops have comparable magnitudes. In the case of the charge
radii, the loops give the dominant contribution to the isovector EM charge
radius while the $c_\pm$ give the dominant contribution to the isoscalar
EM charge radius. It is evident, then, that one cannot rely on either the loop
or the \lq\lq counterterm" contributions alone to account for the nucleon's
EM moments.

	In the case of the strangeness magnetic moment and radius, one would
ideally follow a similar strategy. However, the coefficients $b^s$ and 
$c^s$ are unknown. The reason is that these constants depend on $b_0$ and
$c_0$ as well as the $b_\pm$ and $c_\pm$. Since the baryon number magnetic
moment and charge radius for the octet baryons have not be measured, $b_0$ and
$c_0$ are un-determined. In fact, by virtue of Eq.~(\ref{eq:scur}), 
measurements of
the strangeness radius and magnetic moment of the nucleon would provide
a determination of the corresponding quantities for the baryon number
current and, through Eqs.~(\ref{eq:bcts}-\ref{eq:ccts}), 
would fix $b_0$ and $c_0$. Moreover, given
the situation in the EM case, one would not be safe in assuming that
$b^s$ and $c^s$ differ significantly in magnitude from unity. Indeed, one
has no reason to expect, based on any symmetry principle, that either the
loops or chiral counterterms should give the dominant contributions to the
strangeness radius and magnetic moment. 
Thus, chiral perturbation theory, 
in its purest form, cannot make a prediction for the 
strangeness vector current matrix elements. 

In arriving at this conclusion, we did not include decuplet baryons in
the loops nor the sub-leading non-analytic loop contributions ($\sqrt{m_s}
\ln m_s$ in the case of the strangeness magnetic moment) as was done in
Ref. \cite{Jen93}. In that work, 
it was found that the dominant loop contribution
to the magnetic moments is
${\cal O}(\sqrt{m_s})$ and that the inclusion of the decuplet states
does not have the same kind of effect as in the axial vector matrix elements,
where non-negligible octet-decuplet cancellations occur for the loop 
contributions. Similarly, we did not use the one-loop corrected axial
meson-nucleon couplings. Although from a formal standpoint the difference
between tree-level and one-loop corrected couplings is of higher order than
we are considering here, the authors of Ref. \cite{Jen93} obtained a better fit
to the baryon magnetic moments with the corrected couplings. The use of the
latter effectively reduces the size of the large kaon loop contributions.
As we note below, the physics which modifies one loop results largerly
amounts to kaon rescattering (see, {\em e.g.} Ref. \cite{Fed58}). Employing 
one-loop corrected axial couplings in the one-loop magnetic moment
calculation incorporates some, but not all, rescattering contributions. 
It is not entirely clear that the impact of
two-loop contributions to the magnetic moment
is numerically less significant than the replacement of tree-level with
one-loop corrected axial couplings in the one-loop magnetic moment calculation.
In the present instance, we avoid this issue altogether and restrict
our attention to one-loop effects.

\medskip
\centerline{\bf B. Chiral Models}
\medskip

\def\gvnn{{g_\sst{VNN}}}
\def\fv{{f_\sst{V}}}
\def\mvs{{m_\sst{V}^2}}
\def\mv{{m_\sst{V}}}

The conclusion of the foregoing analysis implies that in order to make 
predictions for the nucleon's strangeness moments, one must go beyond the
framework of CHPT and invoke additional, model-dependent assumptions. To this
end, a number of possibilities present themselves. We consider three such
model-approaches: (a) resonance saturation, (b) kaon cloud dominance, 
and (c) constituent chiral quarks.

\bigskip
\noindent{\bf Resonance saturation.} One could, for example,
attempt to estimate the $b^s$ and $c^s$ by assuming that the corresponding
terms in $\Delta{\cal L}$ arise from 
$t$-channel vector meson exchanges. The rationale
for such an approach derives primarily from one's experience in the
purely mesonic sector where, at ${\cal O}(p^4)$ in the chiral expansion,
one encounters ten scale-dependent counterterms, $L_i^r(\mu)$ 
\cite{Eck89a,Eck89b}. 
Five of these counterterms ($i=1,2,3,9,10$) agree quite well with the 
predictions of vector meson exchange when the renormalization scale is
chosen to be $\mu\approx\mro$. Of particular interest is the pion EM
charge radius, which receives a contribution from $L_9^r(\mu)$. This
counterterm contribution dominates $\langle r_\pi^2\rangle$, with the
one-loop contribution giving roughly 7\% of the total (for $\mu=\mro$).
Were this situation to carry over into the arena of the nucleon's vector
current form factors, one would then expect the counterterms $b^a$ and
$c^a$ to be given by vector meson resonances, as shown in Fig. 2. 

	To explore this possibility further, one requires the couplings
of $J^{PC}=1^{--}$ vector mesons $V$ to spin-$1/2$ baryons and to electroweak
vector bosons. Although it is conventional to describe the vector mesons
by a vector field $V_\mu$, we choose instead to follow 
Refs. \cite{Eck89a,Eck89b} and work
with a formulation in terms of a two-index anti-symmetric tensor, $V_{\mu\nu}$.
This formulation offers the advantages that (a) it is straightforward to write
down a gauge-invariant Lagrangian for the interaction of the vector meson with
electroweak vector bosons, and (b) the contributions from the diagram in Fig.
2 do not affect the normalization of the Dirac form factor at $Q^2=0$. In 
addition, one finds, as shown in Ref. \cite{Eck89a}, that the vector field
formulation does not generate a vector meson contribution to the pion EM
charge radius -- a situation one must remedy by the introduction of an 
additional term at ${\cal O}(p^4)$ in the chiral Lagrangian. No such
term is necessary with the tensor formulation. The primary
cost involved in using the anti-symmetric tensor formulation is the 
presence of a four-index vector meson propagator. For the calculation of
tree-level process such as given in Fig. 2, this cost is not exhorbitant.
Since the details of this formulation and its relation to the vector field
framework are discussed in Refs. \cite{Eck89a,Eck89b}, 
we refer the reader to those papers
and simply give the form of the couplings and results for the nucleon form
factors.  

	The vector meson contributions to the nucleon magnetic moment
and charge radius are generated by the following two $VNN$ effective
Lagrangians:
\begin{equation}
{\cal L}_{VNN}=2G_T\epsilon^{\mu\nu\alpha\beta} v_\alpha
\bbar_v S^v_\beta B_v V_{\mu\nu} + {G_V\over\lamchi}\bbar_v B_v v_\mu
D_\nu V^{\mu\nu}\ \ \ ,
\label{eq:lvec}
\end{equation}
while the gauge invariant coupling vector meson-photon coupling is given by
\begin{equation}
{\cal L}_{V\gamma}={eF_V\lamchi\over\sqrt{2}} V_{\mu\nu} F^{\mu\nu}
\ \ \ .
\label{eq:lvgam}
\end{equation}
A similar expression applies to the coupling of $V$ and $Z$ (the source of
the baryon number current). We have omitted SU(3)-indices for simplicity.
In the formalism of Refs.~\cite{Eck89a,Eck89b}, the field
$V_{\mu\nu}$ has dimension one. The factors of $\lamchi$ have been introduced
to maintain the correct dimensionality while employing dimensionless couplings.
In this respect, our definition of $F_V$ differs from that of 
Refs.~\cite{Eck89a,Eck89b},
where the corresponding coupling has mass dimensions. The values for the
$F_V$ can be extracted from the rates $\Gamma(V\to e^+e^-)$. In the case
of the lightest isovector vector meson, for example, one has $F_\rho=0.132$.

From these couplings and the amplitude associated with the diagram of Fig. 2,
we find the following contributions to the Dirac and Pauli form factors:
\begin{eqnarray}
	F_1^I(Q^2)&=&\sqrt{2} G_V F_V\ {Q^2\over \mvs-Q^2}
\label{eq:foneI} \\
	 F_2^I(Q^2)&=&4\sqrt{2} G_T F_V\ {\mn\lamchi\over\mvs}
	{\mvs\over\mvs-Q^2}\ \ \ , 
\label{eq:ftwoI}
\end{eqnarray}
whered $\mv$ is the vector meson mass.
Again following Ref. \cite{Eck89a}, we have introduced the superscript $I$ to 
denote the results using the tensor formulation (\lq\lq model $I$" in 
Ref. \cite{Eck89a}). Note that $F_1^I(Q^2=0)=0$, so that the vector meson 
resonances do not affect the nucleon charge.
For purposes of comparison, we write down analagous expressions in the
vector field formulation (\lq\lq model $II$"), where one
has $F_1^{II}(0)\not=0$. In this model, one must include an additional
counterterm in the Lagrangian in order to preserve the nucleon's charge.
Alternately, one can work with the difference $F_1^{II}(Q^2)-F_1^{II}(0)$,
which is independent of the nucleon charge. Working with this difference
is equivalent to making a narrow resonance approximation to a subtracted
dispersion relation for $F_1$. Since the value of $F_2(0)$ is
not associated with any conserved charge, no subtraction is required.
The vector field formulation yields, then
\begin{eqnarray}
	F_1^{II}(Q^2)-F_1^{II}(0)&=&{\gvnn\over\fv}{Q^2\over\mvs-Q^2}
\label{eq:foneII}\\
	 F_2^{II}(Q^2)&=&{\gvnn\kappa_\sst{V}\over\fv}{\mvs\over\mvs-Q^2}
	\ \ \ ,
\label{eq:ftwoII}
\end{eqnarray}
where $\gvnn$ and $\kappa_\sst{V}$ give, respectively, the usual vector and
tensor $VNN$ interaction strengths 
\cite{Sak69,Hol89,Gen76,Hoh76,Hoh75} and $\fv$ sets the scale of the
$V\gamma$ transition amplitude. 

	From Eqs. (\ref{eq:foneI}-\ref{eq:ftwoI}) or 
(\ref{eq:foneII}-\ref{eq:ftwoII}), 
one can extract the vector meson contributions to
the nucleon magnetic moment and charge radius and, thus, the corresponding
contributions to the chiral coefficients $b$ and $c$:
\begin{eqnarray}
	b&=&2\sqrt{2} G_T F_V\left({\lamchi\over\mv}\right)^2
	=\left({\gvnn\kappa_\sst{V}\over\fv}\right)
	\left({\lamchi\over 2\mn}\right)
\label{eq:bvec}\\
	c&=&\sqrt{2}G_V F_V\left({\lamchi\over\mv}\right)^2
	=\left({\gvnn\over\fv}\right)
	\left({\lamchi\over\mv}\right)^2\ \ \ ,
\label{eq:cvec}
\end{eqnarray}
where the flavor index \lq\lq $a$" has been omitted for simplicity.
In the case of the nucleon's EM form factors, the expressions in 
Eqs.~(\ref{eq:bvec}-\ref{eq:cvec}),
together with the decay rates for $V\to e^+ e^-$, 
can be used to determine 
the couplings $G_V$, $G_T$, and $F_V$ 
(or $\gvnn$, $\kappa_\sst{V}$, and $\fv$)
\cite{Gen76,Hoh76,Hoh75}. 
Were one also to possess knowledge of the $F_V$ (or $\fv$) associated
with the strangeness matrix elements $\bra{0}\sbar\gamma_\mu s\ket{V}$, one
could then use the expressions in 
Eqs.~(\ref{eq:bvec}-\ref{eq:cvec}) to derive the counterterms for the
nucleon's strangeness form factors. However, one does not at present possess
such knowledge. As a fall-back strategy, one may invoke one's knowledge of the
flavor content of the vector meson wavefunctions, where such knowledge exists.
In doing so, it is useful to follow the spirit of Refs. 
\cite{Jaf89,Gen76,Hoh76} 
and write down dispersion relations for the nucleon form factors:
\begin{eqnarray}
F_1^{(a)}(Q^2)-F_1^{(a)}(0)&=&Q^2\sum_V {a^{(a)}_V\over \mvs-Q^2}
	 +Q^2{\tilde f_1^{(a)} (Q^2)}
\label{eq:foned}\\
	 F_2^{(a)}(Q^2)&=&\sum_V {\mvs b_V^{(a)}\over \mvs-Q^2}+
	 {\tilde f_2^{(a)}(Q^2)}\ \ \ ,
\label{eq:ftwod}
\end{eqnarray}
where the superscript $(a)$ denotes the flavor channel ($T=0,1$ or
strangeness), where the poles arise from vector meson exchange as in Fig. 2,
and where the functions $f_i(Q^2)$ represent contributions from the
multi-meson continuum.\footnote{Note that the continuum contribution need
not enter additively; one may also include it as a multiplicative 
factor \cite{Hoh75,Mer96}. We write 
it additively for simplicity of illustration.} 
In the works of Refs. \cite{Jaf89,Gen76,Hoh76}, the continuum
contributions were neglected in the isoscalar and strangeness channels. 
In the spirit of resonance saturation, we retain the leading, non-analytic
loop contributions as an estimate of the continuum terms and assume
that the counterterms $b^a$ and $c^a$ are dominated by the vector meson
pole contributions. From Eqs.~(\ref{eq:foneI}-\ref{eq:ftwod}), 
these counterterms are easily related to the pole residues:
\begin{eqnarray}
	b^a&=&\left({\lamchi\over 2\mn}\right)\sum_V b_V^{(a)}
\label{eq:bares} \\
	 c^a&=&\sum_V a_V^{(a)}\left({\lamchi\over\mv}\right)^2
	\ \ \ .
\label{eq:cares}
\end{eqnarray}
Thus, for purposes of determining the chiral coefficients $b^a$ and 
$c^a$, it is just as effective to work with the residues in a pole analysis
of the form factors as it is to try and determine the hadronic couplings
$G_V$ $G_T$, and $F_V$ (or $\gvnn$, $\kappa_V$, and $\fv$). 

A determination of the
residues was carried out by the authors of Refs. \cite{Gen76,Hoh76}, 
who employed a 
three pole fit to the isoscalar EM form factors. The poles were identified,
respectively, with the $\omega$, $\phi$, and one higher mass isoscalar
vector meson $V'$ (for an up-date, see Ref. \cite{Mer96}).
The inclusion of at least two poles was needed in order to reproduce the
observed dipole behavior of the isoscalar form factors. The authors found
that a third pole was needed in order to obtain an acceptable $\chi^2$
for the fit. Subsequently, Jaffe \cite{Jaf89} 
observed that since the physical $\omega$
and $\phi$ are nearly pure $u\ubar+d\dbar$ and $s\sbar$ states, respectively,
one can relate the residues appearing in the strangeness form factor
dispersion relations to those associated with the isoscalar EM form factors:
\begin{eqnarray}
{a_\omega^s\over a_\omega^{T=0}}&=&-\sqrt{6}\left[{\sin\epsilon\over
	\sin(\epsilon+\theta_0)}\right]\\
\label{eq:jafa}
	 {a_\phi^s\over a_\phi^{T=0}}&=&-\sqrt{6}\left[{\cos\epsilon\over
	\cos(\epsilon+\theta_0)}\right]\ \ \ ,
\nonumber
\end{eqnarray}
where $\epsilon$ is the mixing angle between the pure $u\ubar+d\dbar$ and
pure $s\sbar$ states and $\theta_0$ is the \lq\lq magic" 
octet-singlet mixing angle giving rise to these pure states.  Analogous
formulae apply for the residues appearing in the expressions for $F_2$.
From Eqs.~(\ref{eq:bares}-\ref{eq:jafa}),
one may now determine the $\omega$ and $\phi$ contributions to the
constants $b^s$ and $c^s$.

A determination of the remaining residues $a_{V'}^s$ and $b_{V'}^s$ is
more problematic. One does not possess sufficient knowledge of the $V'$
flavor content to derive a simple relation between the strangeness and
isoscalar EM residues. One must therefore employ alternative strategies.
Jaffe arrived at values for the  $a_{V'}^s$ and $b_{V'}^s$ by imposing
conditions on the asymptotic behavior of the form factors ($Q^2\to\infty$).
Using a three pole fit, with all masses and two residues fixed, one is
only able to require that $F_1$ vanish as $1/Q^2$ and $F_2$ as $1/Q^4$. 
These asymptotic conditions are more gentle than one would expect based
on the most na\"\i ve quark counting rules. Consistency with the latter would
require the inclusion of more poles with unknown masses and residues than
used in the fits of Refs. \cite{Jaf89,Gen76,Hoh76}. 
Since the adequacy of these quark
counting rules for strangeness form factors is itself not clear,
and since one's predictions for the nucleon's strangeness radius and magnetic
moment within this framework are non-trivially dependent on one's assumptions
about asymptopia, this approach to treating the $V'$ contribution is ambiguous
at best. 

Another alternative is to note that in the fits of Ref. \cite{Hoh76}, the $V'$ 
contributes very little to the isoscalar mean square radius and anomalous
magnetic moment (less than 10\% in the fits with the best $\chi^2$). Indeed,
the primary benefit of including the $V'$ was to obtain acceptable $\chi^2$
over the full range of $Q^2$ used in the fit; it's impact on the value of
the form factors and their slopes at the origin is minimal. The latter
result is not surprising, since the $V'$ contribution to the low-$|Q^2|$
behavior of the form factors is suppressed by powers of $(m_\sst{\omega,
\phi}/m_{V'})^2$ relative to the $\omega$ and $\phi$ contributions. 
It would seem reasonable, then, to neglect the $V'$ when seeking to determine
the leading, non-trivial $Q^2$ behavior of the strangeness form factors. This
strategy is the one we adopt here. The uncertainty associated with neglecting
the $V'$ pole is certainly no greater than the ambiguity 
one encounters when using large $Q^2$ conditions to determine low-momentum
constants. Moreover, in the present study, we seek to make statements only 
about the strangeness radius and magnetic moment, and not about the full
$Q^2$-dependence of the strangeness form factors. Hence,  including additional
poles beyond the $\omega$ and $\phi$ is not a strong necessity. In principle,
the $V'$ or even higher mass isoscalar $1^{--}$ resonances could generate
large contributions to $b^s$ and $c^s$. Such result would be surprising, 
based on the situation in the isoscalar channel. Nevertheless, the latter
possibility cannot be ruled out on general grounds. Indeed, one's lack of
knowledge of the contributions from the $V'$ and beyond constitutes one of
the weak points in the resonance saturation model. To be on the conservative 
side, we choose to omit contributions about which we have no knowledge.

	In Table I, we quote results for the nucleon's Dirac strangeness
radius and magnetic moment assuming the $\omega$ and $\phi$ residues saturate
the constants $b^s$ and $c^s$. We obtain these constants using the Jaffe
relations in Eq. (30), the results from fit 8.2 of Ref. \cite{Hoh76}
(which gives the best
$\chi^2$), and Eqs.~(\ref{eq:bares}-\ref{eq:cares}). 
In viewing these results, a few additional caveats
should be kept in mind. First, the one-loop contributions to the isoscalar
and strangeness radius and magnetic moment quoted above represent a subset
of a larger class of continuum contributions to the moments. In the language
of Eqs.~(\ref{eq:foned}-\ref{eq:ftwod}), 
these loop calculations give an estimate of the two kaon 
intermediate state contributions to the functions $\tilde f(Q^2)$.
Contributions from the $3\pi$, $5\pi$, $KK\pi,\ldots$ states have not been
included. 

	Second, the resonance saturation model is only partially successful
in the case of the nucleon's EM moments, in contrast to the situation 
with the pion form factor. To illustrate, we consider the EM Dirac charge radii.
In the case of the isovector radius, the loop contribution is signficantly
larger than the experimental value: $\rho^{T=1}_\sst{LOOP}/\rho^{T=1}_\sst{
EXPT}\approx 1.5$ (taking $\mu\approx\mro$).\footnote{In the work of
Ref. \cite{Gas88}, 
only $\pi$ loops were considered and the result for
$\rho^{T=1}_\sst{LOOP}$ is closer to the experimental value. Our result
also includes the $K$-loop contribution. Although the calculation of Ref.
\cite{Gas88} was carried out without using the heavy baryon formalism, the 
result agrees with the chiral log of the heavy baryon calculation. The
results for the magnetic moment differ, however.}
One therefore requires a contribution from $c^{T=1}$
which cancels about 40\% of the loop contribution. The $\rho$ meson
contribution to $c^{T=1}$, computed using the values of $\fv$ taken from
$e^+ e^-$ data and $\gvnn$ determined from fits to $NN$ scattering amplitudes
\cite{Hol89}has the wrong sign to bring about this cancellation.
In fact, a careful analysis of the isovector spectral function
for the Dirac form factor, which contains information about both the 
continuum and $\rho$ resonance contributions, can be used to extract a value
for $\gvnn$ consistent with the value used in $NN$ scattering studies
\cite{Hoh75}. Such an analysis includes pion rescattering corrections
which reduce the two pion continuum contribution and allow for a larger
$\rho$ pole term. In the case of the isoscalar Dirac radius, on the 
other hand, the kaon loop contribution to the
isoscalar Dirac radius is about 15\% of the experimental value. To the
extent that the multi-pion continuum and kaon rescattering
contributions are neglegible, 
$\rho_\sst{D}^{T=0}$ is therefore resonance dominated. 

Third, one must note the presence of an issue of consistency between the
resonance saturation model employed here and the ways in which the 
constants $b^s$ and $c^s$ have been extracted. In the analysis of Refs.
\cite{Gen76,Hoh76}, 
no continuum contributions were included in the fit to the isoscalar
form factors. Such an approximation may be valid in the case of the isoscalar
Dirac radius, for which the loop contributions represent a reasonably small
fraction of the total. One might expect, then, that
a fit which included the $2K$ continuum contributions would not yield
residues $a_{\omega,\phi}^{T=0}$ differing appreciably from the values
of Refs. \cite{Gen76,Hoh76}.
Stated differently, not only is the constant $c^{T=0}$ dominated by resonance,
but apparently so is the entire isoscalar Dirac radius. 
Conversely, the ${\cal O}(\sqrt{m_s})$ kaon loop 
contribution to the isoscalar anomalous magnetic moment is large:
$\kappa^{T=0}_\sst{LOOP}/\kappa^{T=0}_\sst{EXPT}\approx 20$. Chiral
perturbation theory therefore requires a large constant $b^{T=0}$ to
cancel most of this loop contribution. Unfortunately, the most reliable
information one has on the resonance contributions to the isoscalar magnetic
form factor is derived from the fits of Ref. \cite{Hoh76}, which included
no continuum. The residues $b_{\omega,\phi}^{T=0}$ obtained from these fits
are small (on the order of $\kappa^{T=0}_\sst{EXPT}$) and, therefore, cannot
cancel the large kaon loop contribution. Presumably, a reanalysis of the
isoscalar magnetic form factor which included the kaon continuum at one-loop
order would yield larger values for the residues. We would conclude from
these observations that, of the resonance saturation predictions quoted in
Table I, the value for the Dirac radius is the more credible.

\medskip
\noindent{\bf Kaon cloud dominance.}
A second possibility is to relax the requirement that one undertake a
consistent chiral expansion and use kaon loops alone to make a prediction. The 
rationale for this approach has a two-fold basis. The first follows from
a geometric interpretation of the
nucleon charge radius, wherein it characterizes a spatial asymmetry in
the charge distribution. In this picture, a spatial polarization of the
strange sea arises from fluctuations of the nucleon into a kaon and
strange baryon. The kaon, having about half the mass of the lightest strange
baryons lives on average further from the nucleon center of mass than
the strange baryon. One would expect, then, to obtain a negative value
for $\langle r_s^2\rangle$ (positive value for $\rho^s$), since the kaon
carries the $\bar s$. Implicit in this picture is an assumption that
$s\bar s$ pair creation by the neutral gauge boson probe, 
which also contributes to the Dirac or electric form
factors and which appears partially in the 
guise of resonance contributions, is negligible compared to the mechanism
of $s\bar s$ spatial polarization. The kaon cloud dominance approach
also assumes that the multi-pion contribution is negligible when compared
to that of the kaon cloud, ostensibly because the pion contains no valence
$s$ or $\bar s$ quarks.

	The second motivation draws on the result of a pion loop
calculation of the nucleon's EM form factors first carried out
by Bethe and DeHoffman \cite{Bet55}. 
This calculation was performed using the equivalent of the linear 
$\sigma$-model. At the time they were reported,
the results were in surprising agreement with the experimental
values for the nucleon's charge radii and magnetic moments, despite the
large value of the $\pi N$ coupling which enters this perturbative calculation.
The lore which developed in the aftermath of this calculation is that
the pion cloud dominates the nucleon's isovector EM moments and that a 
one-loop calculation sufficiently incorporates the physics of the pion
cloud. Were this situation to persist in the strangeness sector, one would
expect that the kaon cloud gives the dominant contribution to the
$\rho^s$ and $\mu^s$ and that a one loop calculation would suffice to
give their correct magnitude and sign.

A variety of one-loop calculations have been performed assuming that the
kaon cloud dominates the strange form factors. For example, 
the authors of Ref. \cite{Mus94b} computed $\rhostr$ and $\mustr$
within the context of the SU(3) linear $\sigma$-model. Within this framework,
the leading strangeness moments are U.V. finite. Nevertheless, the calculation
was performed by including hadronic form factors at the $K N\Lambda$ vertices,
drawing on results of fits to baryon-baryon scattering in the one meson
exchange approximation which find better agreement with data if hadronic form
factors are included. The authors of Refs. \cite{For94,Coh93}
extended this approach to
compute both the leading moments as well as the non-leading
$Q^2$-dependence of the strangeness form factors using a hybrid
kaon-loop/vector-meson pole model. Although the hybrid model goes beyond a
simple one-loop approximation, it nevertheless represents a type of kaon
cloud model inasmuch as non-resonant multi-pion contributions are omitted.
Another variation of this general approach is a study
performed using the cloudy bag model (CBM) and the \lq\lq cloudy" constituent
quark model (CCQM)\cite{Koe92}. The CBM represents a kind
of marriage of the MIT bag model with spontaneously broken chiral symmetry.
The strength of the meson-baryon vertices is determined by the 
meson-quark coupling and the quark's bag model wavefunction.
The CCQM is similar in spirit, though in this case the non-relativistic
constituent quarks are confined with a harmonic oscillator potential.
In effect, the CBM and CCQM 
calculations represent kaon loop calculations in which
the $N\Lambda K$ and $N\Sigma K$ form factors are determined by the dynamics
of the particular models. More recently, Geiger and Isgur have extended 
the kaon cloud idea to include
one loop contributions from all known strange mesons and baryons using the
non-relativistic quark model to obtain a nucleon/strange hadron vertex 
function \cite{Gei96}. 

In all cases, these models include contributions which
are both non-analytic and analytic in the $s$-quark mass, effectively modelling
contributions from the relevant
higher-dimension operators appearing in an effective Lagrangian.
Moreover, in each instance the loop integration was cut-off at some momentum
scale by including form factors at the hadronic vertices. In both respects,
a consistent chiral expansion is lost. In principle, higher-order loop
contributions could yield terms of the same chiral order as
some of the analytic terms retained from the one-loop amplitudes. Similarly,
the use of hadronic form factors with a cut-off parameter breaks the
consistency of the expansion because a new scale is introduced
({\em e.g.}, the $1/\hbox{hadron size}$) and/or because the form factor itself
contributes like an infinite tower of higher-dimension operators.

One might argue that since $m_K/\lamchi$ is not small, the
chiral expansion is not all that useful in the case of strange quarks and
that models which are inconsistent with this expansion may yet be 
credible \cite{Hat90,Mei94}.
We wish to illustrate, nonetheless, that the approach of kaon cloud
dominance still presents a host of uncertainties. To do so, we repeat
the calculation of Ref. \cite{Mus94b}
in the framework of the non-linear SU(3) $\sigma$-model,
corresponding to the meson-baryon Lagrangian of Eq.~(\ref{eq:lbar}). 
In this case,
the strangeness radius is U.V. divergent, unless one includes form factors
at the hadronic vertices. A simple choice, and one which renders the loop
calculation most tractable, is the monopole form:
\begin{equation}
F(k^2)={\mks-\Lambda^2\over k^2-\Lambda^2}\ \ \ ,
\label{eq:mesff}
\end{equation}
where $k$ is the momentum of the kaon appearing at the $KN\Lambda$ vertex
and $\Lambda$ is a momentum cut-off. The monopole form above was employed in
the Bonn potential fits to baryon-baryon scattering, and value of the
cut-off $\Lambda\sim 1.2 - 1.4$ GeV was obtained \cite{Hol89}. 
Various kaon cloud
models differ, in part, through the choice of form for $F(k^2)$ and the
value of the cut-off parameter. 

The inclusion of a hadronic
form factor necessitates the introduction of additional, \lq\lq seagull"
graphs in order to maintain the gauge invariance of the calculation
(Fig. 1c,d). Without
these new graphs, the loop calculation with hadronic form factors does not
satisfy the vector current Ward-Takahashi identity. 
It was shown in Refs. \cite{Mus94b,For94}
that use of the minimal subsitution $k_\mu\to k_\mu+i{\hat Q}Z_\mu$ 
in $F(k^2)$ generates a set
of seagull vertices whose loop graphs restore agreement of the calculation
with the W-T identity. It is straightforward to show that for a meson-nucleon
vertex of the form
\begin{equation}
\mp i F(k^2) k^\lambda{\tilde\Pi}\ \ \ubar\gamma_\lambda\gamma_5 u
\label{eq:vert}
\end{equation}
the corresponding seagull vertex is
\begin{equation}
\{\pm i Z_\mu(Q^\mu\pm 2 k^\mu){[F((Q\pm k)^2)-F(k^2)]\over [(Q\pm k)^2-k^2]}
k^\lambda [\Qhat, {\tilde\Pi}]+i Z^\lambda F((Q\pm k)^2) 
[\Qhat, {\tilde\Pi}]\}\ \ \ubar\gamma_\lambda
\gamma_5 u
\label{eq:seag}
\end{equation}
where ${\tilde\Pi}$ is the pseudoscalar octet matrix defined in Section II, 
$Q_\mu$ is
the momentum of the source $Z$ (for EM or baryon number current), ${\hat Q}$
is the corresponding charge, and where the upper (lower) sign corresponds
to in incoming (outgoing) meson. 

In Section IV and Table I 
we give the results of the kaon loop calculation using
the non-linear SU(3) $\sigma$-model and hadronic form factors 
(as in Eq.~(\ref{eq:mesff})) as a function of the cut-off $\Lambda$. 
We compare these results with those of other kaon cloud models in order
to estimate the range in predictions which arises under the rubric
of kaon cloud dominance. Indeed, the existence of such a range reflects
the ambiguities associated with this general approach.
First, as mentioned previously, one has already abandoned a consistent 
chiral expansion. Consequently, one has no rigorous justification for
retaining only one-loop contributions. Second, the choice of hadronic form 
factor is not unique. In the CBM, for example,
the form of the effective $F(k^2)$ is approximately
Gaussian rather than monopole as used here. Moreover, the scale of
the momentum cut-off is set by the inverse bag radius, which is on
the order of a few hundred MeV \cite{Koe92}. 
The CBM constitutes a chiral model with a different underlying 
physical picture than the non-linear $\sigma$-model, and its parameters can
be tuned to produce agreement with at least some of
the nucleon's EM form factors. One has no strong phenomenological
reason, then, to choose one model -- corresponding to one form for
$F(k^2)$ -- over another; only a model preference. Third, the prescription for
maintaining gauge invariance is also not unique. The one shown above is
a {\em minimal} procedure. One may include additional seagull
contributions which are purely transverse and, therefore, do not affect the
W-T identity. The presence of these additional terms may, nevertheless, 
affect one's results for the form factors. Finally, this approach omits
resonance contributions altogether. This omission by itself ought to raise
concern. Indeed, it has long been known, from dispersion theoretic studies,
that a significant part of the nucleon's isovector EM form factors contain
important resonance contributions \cite{Hoh75}. 

\medskip
\noindent{\bf Constituent quarks.} The final model approach we consider
entails treating the nucleon's strangeness matrix elements as arising from
the strangeness \lq\lq content" of constituent $U$- and $D$-quarks. The
motivation for this approach derives from  
a picture of the constituent quark as a current quark of QCD surrounded
by a sea of gluons and $q\bar q$ pairs. It follows that the nucleon's 
strangeness radius and magnetic moment arise from  the corresponding
quantities for the consituent $U$- and $D$-quarks 
\cite{Kap88}.\footnote{The CCQM calculations of Ref. \cite{Koe92} omit
contributions from the strangeness content of the constituent quarks. 
Only the kaon cloud around the entire bag of quarks is considered.}
The procedure one follows within this framework is essentially 
the quark model analog of the one-body approximation
made in computing nuclear current matrix elements \cite{Ito95}. 
Specifically, one
derives an operator associated with the individual constituents (quarks,
nucleons) and computes a matrix element of that operator using the appropriate
bound state (hadron, nucleus) wavefunction. Chiral symmetry is invoked in
deriving the constituent quark strangeness current operators. Such
a calculation of $\rho^s_\sst{S}$ was performed by the authors of
Ref.~\cite{For94}, using the Nambu/Jona-Lasinio (NJL) model \cite{Nam61}
to compute the constituent quark strangeness radii. 

An alternative is to adopt a chiral quark model framework, wherein the
constituent quark strangeness currents arise from fluctuations of the
$U$- and $D$-quarks into a kaon plus a constituent $S$-quark.
The contributions from the individual $U$- and $D$-quarks are added to
give the total nucleon strangeness matrix element using a quark model
spin-space-flavor wavefunction, as illustrated schematically in Fig. 3. 
The strength of the kaon-constituent quark interaction is governed by the
parameter $\ga$ appearing in the chiral Lagrangian of Eq.~(\ref{eq:lquark}). 
This parameter
can be determined by using the constituent chiral quark model to compute
the nucleon's axial vector current. Since $\ga$ enters the strangeness matrix
element of the nucleon at one-loop order, one need only determine it at
tree level (see below). 

	It is worth noting that the chiral quark model does not suffer from
the same lack of convergence which plagues the na\"\i ve baryon 
chiral Lagrangian
due to the size of $m_B$. Since the constituent quark mass is considerably
smaller than $\lamchi$, one has reason to believe that higher-order corrections
to the leading order Lagrangian in Eq.~(\ref{eq:lquark}), 
as well as higher-order loop
effects, will be suppressed. On the other hand, the ambiguity associated with
the coefficients $b^s$ and $c^s$ remains. In the case of chiral quarks, one
may still write down corrections to ${\cal L}_Q$ associated with the
magnetic moment and Dirac charge radius of a constituent quark that are,
respectively, of lower order and the same order in $1/\lamchi$ as the
corresponding contributions from loops: 
\begin{equation}
\Delta{\cal L}={b^a_q\over 2\lamchi}\psibar\sigma_{\mu\nu}{\hat Q}
	\psi F^{\mu\nu} -{c_q^a\over\lamchis}\psibar\gamma_\mu
	{\hat Q}\psi\partial_\nu F^{\mu\nu}\ \ \ ,
\label{eq:ctquark}
\end{equation}
where $\hat Q$ is the appropriate charge (EM or baryon number),
$F^{\mu\nu}$ is the field strength associated with the corresponding source,
and the \lq\lq $a$" superscript denotes the flavor channel \cite{Expc}. 

As in the case of the baryon chiral Lagrangian, the coefficients 
$b_q^0$ and $c_q^0$ in the SU(3)-singlet channel cannot be determined
from known moments. Consequently, one must invoke additional model
assumptions in order to make chiral quark model predictions for the
nucleon's strangeness matrix elements. In the present study, we adopt
the following strategy. First, we 
simply omit the contributions from the $b_q^a$ and $c_q^a$ and take
the constituent-quark/kaon one-loop contribution as an indication of
the scale of the constituent quark strangeness radius and magnetic moment.
Although this assumption, which represents our model {\em ansatz}, may
appear to be a drastic approximation, it is no more questionable than
would be any attempts to make model predictions for the singlet coefficients
$b_q^0$ and $c_q^0$. Second, we cut the loops off at $\lamchi$, 
effectively restricting the virtual Goldstone bosons
to have momenta less than the
scale of chiral symmetry breaking. An alternative would be to use
dimensional regularization and subtract terms proportional to 
${\cal C}_\infty$ (equivalent to $\overline{\hbox{MS}}$ renormalization).
Since we are interested only in obtaining the scale of the constituent $U$-
and $D$-quark strangeness current and not in making airtight predictions,
either approach would suffice. In order to cut the loops off in a gauge
invariant manner, we employ form factors at the quark-kaon vertices
introducing the appropriate seagull graphs as necessary to preserve the
W-T identities. For simplicity, we use the monople form of Eq.~(\ref{eq:mesff}),
taking the cut-off parameter $\Lambda\sim\lamchi$.  In effect, we repeat
the non-linear $\sigma$-model calculation discussed above for constituent
quarks rather than nucleons. 

	The results of the loop calculation generate effective, constituent
quark strangeness current operators
\begin{equation}
\bra{Q} \sbar\gamma_\mu s\ket{Q}\vert_\sst{LOOP}\rightarrow
{\hat J}^\sst{STRANGE}_\mu={\hat\psibar}\left[F^{(s)}_{1Q}\gamma_\mu + 
{iF^{(2)}_{2Q}\over 2 m_Q}\sigma_{\mu\nu}Q^\nu\right]{\hat\psi}\ \ \ ,
\label{eq:sop}
\end{equation}
where $Q$ denotes a constituent quark and $\hat\psi$ is a constituent
quark field operator. Nucleon matrix elements of ${\hat J}_\mu^\sst{STRANGE}$
may be computed using quark model wavefunctions.
We choose to employ wavefunctions in the light-front formalism, since this
framework allows one to use the on-shell constituent quark current (the form
in Eq.~(\ref{eq:sop}) 
and allows one to perform boosts along the direction of momentum
transfer as needed to properly account for the nucleon's center of mass motion.
Although we are concerned only with the leading, non-trivial $Q^2$-dependence
of the strangeness form factors, it is worth noting that the light-front
quark model has successfully reproduced the the nucleon's EM form
factors over a significant range in momentum-transfer \cite{Chu91,Bro94,Azn82}.
We also follow the authors of 
Ref. \cite{Azn82}, who take a tree-level value for
the meson-quark coupling $g_\sst{A}=1.0$ and an oscillator parameter
$\gamma=1.93$ fm~$^{-1}$ and reproduce the nucleon's isovector axial
charge to within 5\%. 

As in the case of the other model approaches discussed here, one should note
the shortcomings of the chiral quark model. Perhaps the primary difficulty
is a conceptual one involving double counting. As noted in Ref. \cite{Man84}, 
the chiral quark effective theory contains both the pseudoscalar $Q\bar Q$ bound
states as well as the octet of light pseudoscalar Goldstone bosons. To the
extent that the latter are also $Q\bar Q$ bound states, the theory contains
the same set of states in two different guises. The authors of 
Ref. \cite{Man84} argue,
based on a simple Goldstone boson-$Q\bar Q$ 
mixing diagram, that the mass of the bound
state must be either somewhat greater than $\lamchi$, in which case it lies
outside the realm of the low-energy effective theory, or infinity, in which
case it is unphysical. One would conclude that the Goldstone boson
octet is distinct from the lightest $Q\bar Q$ states of the theory. A 
study of meson spectroscopy, however, suggests otherwise. Indeed, the
pattern of mass splittings in the $BB^*$, $DD^*$, $KK^*$ and $\pi\rho$
systems is remarkably consistent with the mass splittings in conventional
quarkonia between the lightest $^3S_1$ and $^1S_0$ $Q\bar Q$ states
\cite{Isg92}.
This pattern strongly suggests that the Goldstone bosons
are the lightest $Q\bar Q$ bounds states of the effective theory \cite{Isgpc}, 
in conflict with the conclusions of Ref. \cite{Man84}. 
There do exist methods for
constructing a chiral quark effective theory which includes mesons as specific
degrees of freedom while avoiding the double counting problem (see, 
{\em e.g.} Ref. \cite{Fra91}).  
However, performing a calculation at this level of
sophistication lies beyond the scope of the present study. Our goal is
simply to illustrate how predictions for $\rho^s$ and $\mu^s$ compare 
between models where chiral symmetry is invoked at a microscopic level
and those in which it is included at the purely hadronic level.

\section{Results and Discussion}

In this section, we give predictions for the nucleon's strangeness radius and 
magnetic moment using the three chiral model approaches discussed above. 
These results
are summarized in Table I, where we also give predictions from four previously
reported approaches sharing some elements in common with those discussed here.
For illustrative purposes, we also display in Fig. 4 the dependence of
$\rhostr$ and $\mustr$ on the hadronic form factor cut-off parameter and 
pseudoscalar meson mass entering the non-linear $\sigma$-model calculation.

\vfil
$$\hbox{\vbox{\offinterlineskip
\def\strut{\hbox{\vrule height 15pt depth 10pt width 0pt}}
\hrule
\halign{
\strut\vrule#\tabskip 0.2cm&
\hfil$#$\hfil&
\vrule#&
\hfil$#$\hfil&
\vrule#&
\hfil$#$\hfil&
\vrule#&
\hfil$#$\hfil&
\vrule#\tabskip 0.0in\cr
& \multispan7{\hfil\bf TABLE I\hfil} & \cr\noalign{\hrule}
& \hbox{Model} && \rho^s_\sst{D} && \mu^s && \rho^s_\sst{S}
& \cr\noalign{\hrule}
& \hbox{Resonance Sat.}^{(a)}&& -3.62\ (1.52)&& 1.85\ (2.2) && -5.47\ 
(-0.68)&\cr
& \hbox{NL$\Sigma$M/FF}^{(b)}&& 0.11 && -0.25 && 0.36 &\cr
& \hbox{Chiral Quarks}^{(c)} && 0.53 && -0.09 && 0.62 & \cr
\noalign{\hrule}
& \hbox{Poles}^{(d)}&& -2.43\pm 1.0&& -0.31\pm 0.009 && -2.12\pm 1.0 &\cr
& \hbox{L$\Sigma$M/FF}^{(e)} && 0.1 &&-(0.31-0.40) && 0.41-0.49 &\cr
& \hbox{Hybrid}^{(f)}&& 0.37 && -(0.24-0.32) && 0.61-0.68 & \cr
& \hbox{CBM}^{(g)}&& 0.15 && -0.09 && 0.24 & \cr
\noalign{\hrule}}}}$$
{\noindent\narrower {\bf Table I.} \quad Theoretical predictions for nucleon
strange quark vector current form factors. Columns two and three give
dimensionless mean square Dirac strangeness radius and strangeness anomalous
magnetic moment, respectively. 
The fourth column gives the Sachs strangeness radius:
$\rho^s_\sst{S}=\rho^s_\sst{D}-\mu^s$. To convert to $\langle r_s^2\rangle$,
multiply $\rho^s$ by -0.066 fm$^2$. First three lines give predictions of
chiral models discussed in this work:  
(a) heavy baryon CHPT/resonance saturation employing $\omega$
and $\phi$ residues of Fit. 8.2 of Ref. 
\cite{Hoh76}; numbers in parentheses give
loop contribution for $\mu=m_\rho$; (b) non-linear $\sigma$-model
with hadronic form factors using cutoff mass $\Lambda=1.2$ GeV; (c) chiral
quark model with cutoff mass $\Lambda=1.0$ GeV, oscillator parameter 
$\gamma=1.93$ fm~$^{-1}$,
and $m_U=m_D=0.33$ GeV. Last four lines give previously reported
predictions: (d) three pole model of Ref. \cite{Jaf89};
(e) linear $\sigma$-model of Ref. \cite{Mus94b};
(f) hybrid pole/loop model of Ref. \cite{Coh93}; (g) cloudy bag model of
Ref. \cite{Koe92}.\smallskip}

When viewed from the most \lq\lq impressionistic" perspective, the results
in Table I illustrate the wide spread in predictions one encounters among
approaches relying on chiral symmetry. Indeed, the strangeness radius and
magnetic moment can vary by an order of magnitude and by sign. One ought
to conclude that chiral symmetry by itself is not a terribly restrictive input
principle when it comes to predicting nucleon strangeness. The reason is
essentially that the quantity one wishes to predict is the very quantity
one needs in order to make a prediction 
 -- the SU(3)-singlet vector current moments. In the absence of experimental
information on the latter, the range in one's predictions can be as wide
as the breadth of one's space of chiral models. From the standpoint of 
hadron structure theory,
this situation is not very satisfying, since one would like to possess a
reliable effective-theory framework for interpreting the up-coming 
measurements of the low-energy properities of the $s\bar s$ sea. 

Nevertheless, one may still ask whether there exist reasons to give greater
credibility to one approach over the others. In this respect, we
admit to some bias in favor of the resonance saturation prediction
for the Dirac strangeness radius -- if only because it appears to suffer from
fewer ambiguities than the other approaches. The reasons for this bias can
be summarized as follows: 

\medskip
\noindent (a) Detailed analyses of the isovector charge
radius imply that it is dominated by the lowest continuum state (two pions)
and the lightest isovector $1^{--}$ resonance. 
Although a simple one pion loop
plus $\rho$ resonance calculation over-predicts the isovector charge radius
by a factor of two \cite{Mustbp}, 
the sign and order of magnitude are given correctly.
The over-prediction appears to result from the omission of non-resonant 
pion rescattering corrections \cite{Fed58}. 

\noindent (b) One would expect a similar
situation to persist in the isoscalar and strangeness channels, where the
lightest continuum states are the $3\pi$, $5\pi$, $7\pi$, and $2K$ states
and the lightest isoscalar $1^{--}$ vector mesons are the $\omega$ and
$\phi$. Some thought about chiral counting suggests that the multi-pion
contributions ought to be suppressed relative to the $2K$ contribution. 
To the extent that this suppression holds and that the total, 
non-resonant $2K$ contribution to the isoscalar charge radius 
is at least as small as given the result of the one-loop heavy 
baryon calculation, the isoscalar charge radius would then be dominated by
the lightest isoscalar vector mesons, rendering the fit of Ref. \cite{Hoh76}
quite valid.  The results of this fit indicate that the $\omega$ and
$\phi$ residues dominate the isoscalar Dirac radius; the contribution from
higher mass vector mesons is negligible. Thus, the isoscalar constants 
$b^{T=0}$ and $c^{T=0}$ should be given quite reliably by the 
$\omega$ and $\phi$ contributions. 

\noindent (c) Knowledge of the $\omega$ and
$\phi$ flavor content allows one to translate the 
$\omega$ and $\phi$ contributions to the isoscalar constants into the
corresponding contributions to the strangeness constants, $b^s$ and
$c^s$. If the non-resonant multi-pion contributions to the strangeness
radius are suppressed with respect to the $2K$ contribution, if the
kaon loop contribution (Eq.~(\ref{eq:rsloop})) 
accurately reflects the scale of the two
kaon continuum, and if there are no important vector meson effects beyond
those of the $\omega$ and $\phi$, then $\rho^s_\sst{D}$ ought to be given
accurately by resonance saturation model. 
\medskip

\noindent One should note that this line of
argument avoids the problematic use of assumptions about the strangeness form
factors' large $Q^2$ behavior while incorporating the consistency of the
heavy baryon chiral expansion. 

The logic of this reasoning could break down in a number of ways. Were
a careful dispersion analysis of the isoscalar form factors to reveal large
multi-pion or kaon rescattering contributions, the pole fit of Ref.
\cite{Hoh76}
would need to be re-done, presumably leading to modified values for the
vector meson residues. Moreover, such a situation would imply the presence
of important non-resonant multi-pion and kaon rescattering contributions to the
strangeness form factors as well. In addition, one cannot 
{\em a priori} rule out important contributions from higher mass intermediate
states, even though there is little evidence for such contributions in the
isovector and isoscalar charge radii. 

What about the strangeness magnetic moment? In this case, the resonance
saturation prediction is highly questionable. The reason is that the heavy
baryon calculation gives a large loop contribution to $\kappa^{T=0}$. 
Were this result to accurately reflect a large non-resonant kaon continuum
contribution, the pole fit of Ref. \cite{Hoh76} would be invalid and derived
constants $b^{T=0}$ and $b^s$ un-reliable. On the other hand, 
since $m_K/\lamchi$ is not small, it would not be surprising to find important
kaon rescattering corrections which could cancel the leading term. 
In short, the magnitude and sign of the continuum contributions to
$\kappa^{T=0}$ is uncertain. This uncertainty in turn 
raises doubts about the resonance
saturation prediction for $\mu^s$.
One should also keep in mind that the questionability of this prediction 
for $\mu^s$ implies that the prediction for the Sachs radius is 
also questionable, since $\rho^s_\sst{S}=\rho^s_\sst{D}-\mu^s$.

\medskip
\noindent{\bf Other model uncertainties.}
The remaining model predictions listed in Table I carry as large, 
if not a greater
degree of ambiguity than the resonance saturation prediction for $\mu^s$.

\medskip
\noindent 1.{\sl Kaon cloud dominance.}
Kaon cloud dominance models can be challenged on at least
three grounds: (a) They sacrifice consistent
chiral counting in order to make predictions. This sacrifice comes
about through the introduction of form factors at the hadronic vertices
and through the retention of terms in 
the one-loop amplitudes which are both
analytic and non-analytic in the strange quark mass. Terms of the former
class are indistinguishable from contributions arising from higher-dimension
operators in the chiral Lagrangian. Moreover, higher-order loop graphs
may yield analytic terms of the same chiral order as some of the analytic terms
retained from the one-loop graphs. A consistent chiral approach would 
require the inclusion of all analytic contributions of a given order in
$1/\lamchi$. (b) They do not include 
resonant $t$-channel kaon rescattering or multi-pion contributions (both
resonant and non-resonant). (c) As noted previously,
one encounters ambiguities associated with one's choice of form for the 
hadronic form factors, the size of the cut-off parameter, and the transverse 
part of the covariantizing seagull graphs.

It is instructive to try and quantify the uncertainty associated with these
ambiguities. This effort is most easily accomplished for those of type (c) by
considering the $\Lambda$-dependence of the linear and non-linear 
$\sigma$-model predictions and by comparing these predictions with those of
the CBM and CCQM calculations. Turning first to the issue of the cut-off
dependence,  one may argue about which value of $\Lambda$ to use. 
The results quoted in Table I and Ref. \cite{Mus94b} for the linear
$\sigma$-model were obtained using the Bonn value, 
$\Lambda\sim\Lambda_\sst{BONN}\sim 1.2$ GeV. According to the fits of
Ref. \cite{Hol89}, taking $\Lambda\sim\Lambda_\sst{BONN}$ optimizes
agreement with baryon-baryon scattering data in the one meson-exchange
approximation. For this choice of $\Lambda$, 
however, the corresponding pion loop
contributions to the EM moments are in serious disagreement with the 
experimental values. In fact, there exists no value of $\Lambda$ which
produces agreement between experiment and the linear $\sigma$-model
values for the EM moments. The best choice occurs for $\Lambda\approx 5$
GeV. In this case, experiment and the linear $\sigma$-model agree for
$\kappa^{T=1}$ while the prediction for $\rho^{T=1}$ is 60\% of the
experimental value. Changing $\Lambda$ from $\Lambda_\sst{BONN}$ to
$\Lambda\approx 5$ GeV doubles the prediction for $\mu^s$ and reduces
the prediction for $\rho^s_\sst{D}$ by 25\%. 

The choice of $\Lambda$ in the case of the non-linear $\sigma$-model is 
equally debatable, as a study of the pion-loop contribution to the 
isovector magnetic moment illustrates.\footnote{One would not expect the
pion loop graphs to produce agreement with the isoscalar moments. As the
heavy baryon calculation illustrates, the leading contribution arises from
the diagrams where the current is inserted in the meson line. In the
case of the pion loops, these diagrams only contribute to the isovector
moments.} We find no value of the cut-off
which reproduces the experimental value. 
Choosing $\Lambda\sim\lamchi\sim\Lambda_\sst{BONN}$ 
yields, for example, $\kappa^{T=1}_\sst{LOOP}/
\kappa^{T=1}_\sst{EXPT}\approx 25\%$ ( the
corresponding ratio for the isovector Dirac radius is$\rho^{T=1}_\sst{LOOP}/
\rho^{T=1}_\sst{EXPT}\approx 27\%$). Taking the limit $\Lambda\to\infty$
gives $\kappa^{T=1}_\sst{LOOP}/\kappa^{T=1}_\sst{EXPT}\approx 54\%$
(the isovector Dirac radius diverges in this limit). 
One might argue, then, that choosing any value of the cut-off in the range 
$\lamchi\leq\Lambda\leq\infty$ would be equally justified -- at least for
the magnetic moments which are U.V. finite. As the cutoff is varied over
this range , $\mu^s$ varies from the value quoted in Table I to 
$\mu^s=-1.31$. 
This situation appears to persist in
the case of the CBM and CCQM models as well. The authors of
Ref. \cite{Koe92} originally found values for the cut-off which optimized
agreement with all the nucleon's EM moments. However, a subsequent inclusion of
covariantizing seagull graphs changed the magnetic moment predictions
by 50\% so that an optimum cutoff no longer exists. Consequently, 
the corresponding predictions for the strangeness moments contains an
uncertainty associated with the value of the cut-off. 
 
A comparison between different kaon cloud models reveals a similar degree
of ambiguity. We consider first the linear and non-linear $\sigma$-models.
When one of the baryons is off shell,
as is the case in a loop calculation, the structure of the meson-baryon
vertices differ in the two models, even though the same monopole form
factor was used in both calculations. When one takes $\Lambda\sim\lamchi
\sim\Lambda_\sst{BONN}$, the two models give nearly identical predictions
for $\rho^s_\sst{D}$. One might not be surprised by this result, since 
in both cases the radius contains a chiral log. At least in the chiral
limit, this infrared singularity dominates over contributions analytic 
in $m_K$, and it is essentially terms of the latter type which would be
responsible for any differences in the two predictions.
For $\Lambda\to\infty$, on the other hand, 
$\rho^s_\sst{D}$ diverges in the non-linear
$\sigma$-model but only doubles in value in the linear $\sigma$-model.
In the case of the strangeness magnetic moments,
which contain no infrared or ultraviolate singularities, the model predictions
differ by a factor of about 1.5 for 
$\Lambda\sim\lamchi$\footnote{The lower value for $|\mu^s|$ in the case 
of the linear $\sigma$-model corresponds to $\Lambda=1.2$ GeV.}
but come into closer agreement for $\Lambda\to\infty$. Comparing the
CBM and $\sigma$-model ($\Lambda=\Lambda_\sst{BONN}$) predictions, one
finds CBM gives a 50\% larger Dirac radius but a value for
$\mu^s$ that is a factor of three or four smaller than the linear
$\sigma$-model prediction. 

These comparisons are not definitive. Nevertheless, they suggest a scale for
uncertainty in the kaon cloud dominance predictions that amounts to about
a factor of five or more times the smallest values for $|\rho^s_\sst{D}|$ and
$|\mu^s|$. This rather large model-spread reflects two of the weaknesses
of the kaon cloud dominance approach: (a) the lack of a systematic
expansion procedure ({\em e.g.}, perturbative coupling, chiral, 
$1/N_C$ {\em etc.})
with which to control the one-loop approximation, and (b) the omission
of potentially important short-distance contributions ({\em e.g.}, the chiral
counterterms $(b_a, c_a)$, $t$-channel resonances, {\em etc.}). 

\medskip
\noindent 2.{\sl Vector meson dominance.} Pure vector meson dominance 
models, such as the 
three pole model of Ref. \cite{Jaf89}, omit all non-resonant Goldstone boson
continuum contributions. This practice is not justified by any deep
theoretical arguments, but rather by one's experience in the isoscalar
channel where an acceptable $\chi^2$ is obtained with a three-pole only
fit. The one-loop heavy baryon calculation implies, however, that
the continuum contributions need not be negligible in the strangeness sector.
Moreover, the prediction of Ref. \cite{Jaf89} relies
on questionable assumptions about the asymptotic behavior of the
form factors. The hybrid model of Refs. \cite{For94,Coh93} 
attempts to model both resonant and non-resonant kaon cloud
contributions in a self-consistent manner while
avoiding the problematic assumptions regarding asymptopia. In treating
the two kaon continuum, however, the hybrid model still invokes hadronic 
meson-baryon form factors to cut-off the loop integrals form momenta
above $\lamchi$. Consequently, this approach contains the same ambiguities 
discussed in the preceeding paragraphs.

\medskip
\noindent 3.{\sl Constituent quarks.} Models wherein the nucleon strangeness
moments arise from the strangeness \lq\lq content" of consitutent quarks can
be challenged on three fronts: (a) the credibility of the model itself for
the problem at hand; (b) the procedure for computing the constituent
quark strangeness vector current; and (c) the proper treatment of
collective or many-quark contributions.
The chiral quark model calculation illustrates all three issues. 
As discussed earlier, the problem of double
counting the lightest pseudscalar states raises questions about the na\"\i ve
chiral quark model. Although this difficutly can be addressed using more
sophisticated treatments \cite{Fra91}, the second issue is more problematic.
As in the case of baryon effective theories, the calculation of
the constituent quark strangeness matrix elements employs loops. A consistent
chiral quark calculation implies the presence of terms in the chiral quark
Lagrangian which carry unknown coefficients. These coefficients cannot
be determined without knowing the strangeness matrix elements themselves. 
This dilemma is the same one which hampers heavy baryon 
CHPT as a predictive tool. 
As a fall back, one can employ form factors to cut off the  loop integrals at 
a scale $\lamchi$ as we did in arriving at the numbers in Table I, 
but price one pays is the presence of all the ambiguities encountered when
using hadronic form factors in hadronic loops. Presumably, the spread
in predictions for the constituent quark strangeness currents is as broad
as are the kaon cloud dominance predictions for the nucleon's strangeness
moments. When one considers models other than the chiral quark model, 
the situation is even less controlled. For example, the NJL model, which
provides an alternative model for the constituent quark strangeness radius,
gives a value for $\rho^s_\sst{S}$ that has the opposite sign from the
chiral quark prediction and about 40\% of the magnitude. 

Finally, when one
compares the predictions for chiral quark model and non-linear $\sigma$-model
predictions for $\rho^s_\sst{D}$, one finds that the former is a factor of
five larger than the latter. In both cases, the form for the meson-fermion
vertex is the same, including the approximate
value of the form factor cut-off. One 
faces the question as to whether these two calculations give independent
contributions which should be added, or whether there is some overlap 
between the two. One might argue in favor of the second possibility by
noting that at the quark level, a loop involving a kaon and strange baryon
intermediate state contains diagrams in which a constituent quark fluctuates
into a freely propagating $S$-quark and kaon plus others in which the 
$S$-quark interacts with the other intermediate-state constituent quarks.
In this line of reasoning, the chiral quark model calculation would give
an over-prediction for $\rho^s_\sst{D}$ due to the omission of quark-quark
interactions (many-quark effects). 

\bigskip
\noindent{\bf Experimental Implications}.
With these caveats in mind, it is interesting to ask what the up-coming
neutral current experiments will be able to say about any of the model
predictions in Table I. To that end, some observations about their magnitude
and sign is in order.
In the case of $\mu^s$, all of the models, except for the 
resonance saturation model, give $\mu^s\sim -\mu^{T=0}$. The large
positive value for $\mu^s$ in the resonance saturation model results from 
a large loop contribution -- enhanced by a factor of $\pi(m_K/\lamchi)$ -- and
the questionable absence of correspondingly large vector meson terms.
As for the strangeness radius, all of the calculations containing kaon loops
(except resonance saturation) give the same sign for $\rho^s$ and magnitudes
which vary by a factor of five. The sign corresponds to the na\"\i ve 
expectation derived from the kaon cloud picture in which the
kaon, containing the $\bar s$-quark, lives farther, on average, from the
nucleon's center of mass than does the $\Lambda$, where the $s$-quark resides.
The pole and resonance saturation models, however, give 
strangeness radii having much larger magnitudes and 
the opposite sign. The latter results follow
from the fits of Refs. \cite{Gen76,Hoh76}
which yield a large $\phi NN$ coupling. Note
that in the case of the resonance saturation model, the large $\phi$-pole
contribution to $\rho^s_\sst{D}$ is cancelled to some extent by the
continuum term (loops), whereas in the pure pole model, the $\phi$ contribution
is cancelled to an even greater degree by the questionable $V'$ residue.
An experimental result consistent with resonance saturation value 
$\rho^s_\sst{D}$ would suggest that resonant $t$-channel kaon and multi-pion
rescattering (poles) are the most important physics behind the strangeneness
radius. A significantly smaller result, or one having the opposite sign, 
would imply the presence of large continuum contributions going beyond
one-loop order or important higher-mass resonance terms. 

	How might the various parity-violating electron scattering
experiments do in terms of sorting out among these scenarios? 
The SAMPLE experiment at MIT-Bates \cite{Mck89,Pit94} and
\lq\lq $G^0$" experiment planned for CEBAF \cite{Bec91} 
anticipate a determination of
$\mu^s$ with an error bar of $\pm 0.2$. At this level of precision, 
these experiments could 
confirm the presence of a large strangeness magnetic moment (on the order
of the resonance saturation prediction) or rule out the remaining predictions.
It would be difficult for the
SAMPLE and $G^0$ measurements to confirm any of these remaining predictions
without significantly better precision. As far as the strangeness radius
is concerned, one anticipates a determination of $\rho^s_\sst{S}$ with
an error of $\approx\pm 1.0$ 
from the Hall A and C experiments at CEBAF
\cite{Bec91,Fin91}. These experiments could see a strangeness radius at the
level of the pole and resonance saturation predictions and, at best could
rule out (but not confirm) the remaining entries in Table I. 

Parenthetically, one might note
that forward angle parity-violating electron scattering experiments with
a proton target \cite{Bec91,Fin91,Har93}
are sensitive to the linear combination $\rho^s_\sst{S}
+\mu^p\mu^s$ where $\mu^p\approx 2.79$ is the proton's magentic 
moment \cite{Mus94a,Mus92}. It has been suggested that a determination of this
linear combination is useful as a \lq\lq first pass" probe of the
nucleon's strangeness vector current. Na\"\i vely, a one might conclude
that a small result for this quantity would indicate small magnitudes for
$\GES$ and $\GMS$. In the case of the resonance saturation model, for example,
prediction for $\rho^s_\sst{S}+\mu^p\mu^s$ is an order of magnitude smaller
than the predicted values of either $\rho^s_\sst{S}$ or 
$\mu^s$ alone, owing
to a cancellation between the two terms. Similarly serious cancellations occur
in the case of several of the other model predictions. One ought to be
cautious, therefore, about drawing strong conclusions from a forward 
angle measurement alone. A pair of forward and backward angle measurements,
allowing for separate determinations of $\mu^s$ and $\rho^s$, would be
more relevant to the comparison of model predictions.

\section{Summary}
\label{sec:summary}

In this study, we sought to delineate the extent to which chiral symmetry
can be used to arrive at predictions for the nucleon's strangeness vector
current form factors. Since CHPT has proven quite
useful in other contexts, it is timely to analyze its usefulness in the
case of nucleon strangeness. Moreover, since the role of the $s\bar s$ sea in
the nucleon's low-energy properties is of considerable interest to the hadron
structure community, 
and since significant experimental effort is being devoted to
measuring the nucleon's strange quark form factors, one would like to possess
an effective theory
framework in which to understand the strong interaction dynamics behind
the numbers to be extracted. Ideally, CHPT would have provided such a 
framework. We hope to have convinced the reader that 
nucleon strangeness (or, equivalently, the SU(3)-singlet channel) 
presents barriers to the
applicability of chiral symmetry not present in other cases where symmetry
has proven more useful. To reiterate: the reason for this difficulty is
that the quantity one wishes to predict -- the strangeness (or SU(3)-singlet)
vector current matrix element -- is the same quantity one needs to know in
order to make a prediction. 

Consequently, we turned our attention to chiral models.
We explored three such model approaches as a representative
sampling: a resonance saturation model for the unknown low-energy
constants arising in CHPT; kaon cloud dominance models; and models
in which chiral symmetry is used to obtain the strangeness currents of
constituent $U$- and $D$-quarks. These different approaches yield a rather
broad range of predictions for the nucleon's strangeness radius and magnetic
moment. This situation is not surprising, since none of the approaches
relies solely on the underlying symmetries of low-energy QCD.
We have tried to argue that there may exist one case in which a
chiral model gives a credible prediction: the resonance saturation value for
the Dirac strangeness radius. Of the models considered here, resonance
saturation stays closest to the framework of CHPT while relying on well-defined
phenomenological input. Moreover, it is clear which physics has not been
included (higher-mass poles as well as non-resonant kaon rescattering and 
multi-pion continuum contributions). Should the experimental result for
$\rho^s_\sst{D}$ differ significantly from the resonance saturation value,
one has a reasonable idea of what physics is likely to be responsible for
the discrepancy. In all other cases, including the resonance saturation
prediction for $\mu^s$, we have pointed out what
appear to be good reasons to question the believability of 
chiral model calculations.

Nucleon strangeness remains a highly interesting subject on which 
experiment will shed some light. Assuming that the strangeness radius and
magnetic moment are separately determined with the precision anticipated
for the parity-violation experiments 
\cite{Mck89,Pit94,Bec91,Bei91,Fin91,Har93}, one could, for example, 
test the resonance saturation prediction for $\rho^s_\sst{D}$. 
Given the ambiguities present in the other approaches, an exprimental
result for $\rho^s_\sst{D}$ consistent with the corresponding predictions
would not be conclusive. A similar statement applies to all of the
model predictions for $\mu^s$. In short, 
need for a better theoretical understanding of the nucleon's strangeness
form factors  remains an open problem. In this respect, a more detailed 
analysis within the context of dispersion relations appears to be a promising 
direction. In particular, one ought to look more carefully
at contributions from non-resonant kaon rescattering and multi-pion
intermediate states which are not included in the chiral model approaches
discussed here.
 
\bigskip
\centerline{\bf Acknowledgements}
\bigskip
We wish to thank J.L. Goity, R.L. Jaffe, X. Ji, D. Leinweber, 
U. Meissner, and N. Isgur for useful
discussions and T. Schaefer for raising the question about the resonance
saturation model for the chiral counterterms. We also thank M. 
Frank for a critical reading of the manuscript. 
This work is supported in part by funds 
provided by the U. S. Department of Energy (D.O.E.) under contracts
\#DE-AC05-84ER40150, \#DE-FG06-90ER40561, and \#DE-FG02-95-ER40907.
M.J.M. was also
supported by the National Science Foundation National Young Investigator
Program.

\begin{figure}
\caption{One kaon loop contributions to strangeness vector current
form factors of a non-strange fermion $f$ (nucleon or constituent quark).
Here, $\times$ denotes insertion of the current $\sbar\gamma_\mu s$ and
$f'$ denotes a strangeness $+1$ fermion ({\em e.g.} $\Lambda$ or 
consituent $S$-quark.}
\label{fig: Fig1}

\bigskip
\caption{Resonance contribution to nucleon vector current form factors.
Here $V$ denotes a vector meson and $\times$ denotes a vector current
(EM, strangeness, baryon number, {\em etc.}).}
\label{fig: Fig2}

\bigskip
\caption{Chiral quark model for nucleon strangeness. Shaded circle
represents strange-quark vector current matrix element of a constituent
$U$- or $D$-quark, generated by the processes shown in Fig. 1.}
\label{fig: Fig3}

\bigskip
\caption{Nucleon strangeness vector current moments in the non-linear
$\sigma$-model with hadronic form factors. Dimensionless 
Dirac strangeness radius (panel
(a)) and strangeness magnetic moment (panel (b)) are shown as functions of
the form factor cut-off parameter. To set the scale, note that the nucleon's
dimensionless isovector EM Dirac radius is $\rho^{T=1}_\sst{D}=-4.68$.}
\label{fig: Fig4}
\end{figure}


\begin{references}
\bibitem{Expa} In Ref. \cite{Jen93}, 
	a normalization to $\mn$ rather than $\lamchi$
	was employed for the EM anomalous magnetic moment Lagrangian
and the notation $\mu_D$ and $\mu_F$, rather than $b_+$ and $b_-$, was used
for the coefficients of the commutator and anti-commutator terms,
respectively.
\bibitem{Expb}  In the analysis of Ref. \cite{Jen93}, a fit to the
entire baryon octet was performed. For illustrative purposes, however, we
need consider only the nucleons. In the case of the charge radii, there
exist no measurements for the strangeness $\not= 0$ octet baryons.
\bibitem{Expc} One
may wish to compare the expression in Eq. (\ref{eq:ctquark}) 
with the magnetic moment
term in the chiral quark Lagrangian of Ref. \cite{Man84}. The latter contains
an additional factor of $m/\lamchi$ in order to make explicit the 
chirality-violating character of the magnetic moment interaction. We choose
instead a normalization which goes over to the form of Eq. (\ref{eq:lheav}) 
in the heavy quark limit. For simplicity, we have also ignored
isospin-violating corrections which break the constituent quark EM magnetic
moment relation $\mu_\sst{U}=-2\mu_\sst{D}$.
\end{references}
\end{document}